\documentclass[twocolumn,pre
]{revtex4-1}
\usepackage{color}
\usepackage{graphicx}
\usepackage{dcolumn}
\usepackage{bm}
\usepackage{amsmath}
\usepackage{amssymb}
\usepackage[utf8]{inputenc}
\usepackage[T1]{fontenc}
\usepackage{mathptmx}
\usepackage[percent]{overpic}
\newcommand{\Ron}{R_{\text{on}}}                     
\newcommand{\Roff}{R_{\text{off}}}

\begin{document}
\title{
Global minimization via classical tunneling assisted by collective force field formation
}

\author{$\ ^{\dagger}$F. Caravelli}
\author{$\ ^{\dagger,*}$F. C. Sheldon}
\affiliation{$^\dagger$Theoretical Division (T4) and $^*$Center for Nonlinear Studies,
Los Alamos National Laboratory, Los Alamos, New Mexico 87545, USA}
\author{F. L. Traversa}
\affiliation{MemComputing, Inc. San Diego, California, 92121}

\begin{abstract}
Simple dynamical models can produce intricate behaviors in large networks.  These behaviors can often be observed in a wide variety of physical systems captured by the network of interactions.  Here we describe a phenomenon where the increase of dimensions self-consistently generates a force field due to dynamical instabilities. This can be understood as an unstable (``rumbling") tunneling mechanism between minima in an effective potential. We dub this collective and nonperturbative effect a ``Lyapunov force" which steers the system towards the global minimum of the potential function, even if the full system has a constellation of equilibrium points growing exponentially with the system size. The system we study has a simple mapping to a flow network, equivalent to current-driven memristors. The mechanism is appealing for its physical relevance in nanoscale physics, and to possible applications in optimization, novel Monte Carlo schemes and machine learning. 

\end{abstract}
\maketitle

\textit{Introduction.} 
The dynamics of simple elements interacting in networks can give rise to \emph{emergent} behaviors, which are not exhibited in the dynamics of a single element. These behaviors are inherently multi-element in nature and yet may admit low-dimensional analytical descriptions in terms of coarse-grained variables.  Examples, such as synchronization and phase transitions are often widely applicable, as many different systems may reduce to the same effective model.  In this work we demonstrate such a behavior in a model of input-output networks inspired by memristors.  When weakly driven, the system is described by movement in an effective potential, but at higher driving, instabilities will cause escapes from local minima of the potential which can be captured as a tunneling phenomenon.

Tunneling and escape phenomena are a remarkable feature of quantum mechanical and thermal systems \cite{Sakurai,Kramers}. Thermal and quantum tunneling can occur when a barrier separates two possible minima, and a particle can escape (tunnel) from one minimum to the other aided by either thermal fluctuations or spreading of the wavefunction.
If the barrier is not infinitely high, both thermal and quantum systems have a non-zero probability to tunnel between minima, with a probability exponentially dependent on the height or width of the barrier.
The discovery of quantum tunneling was a radical paradigmatic shift from classical mechanics and underlies many important physical phenomena and technologies (\emph{e.g.} alpha decay, scanning-tunneling microscopy). A similar behavior occurs in non-equilibrium stochastic dynamical systems. Particles at a certain temperature can in fact escape from a metastable state after a typical time proportional to the inverse of the Kramers' rate \cite{Kramers}. Intuitively, thermal fluctuations provide sufficient energy for the particle to escape the local minimum (e.g. a ``kick"). An analogy between these behaviors is not surprising if one considers that the Schr\"{o}dinger equation can be mapped to a non-Markovian stochastic system \cite{Nelson}, or to a classical particle in a non-local force field~\cite{Bohm}. 


\begin{figure}
    \includegraphics[width = \columnwidth]{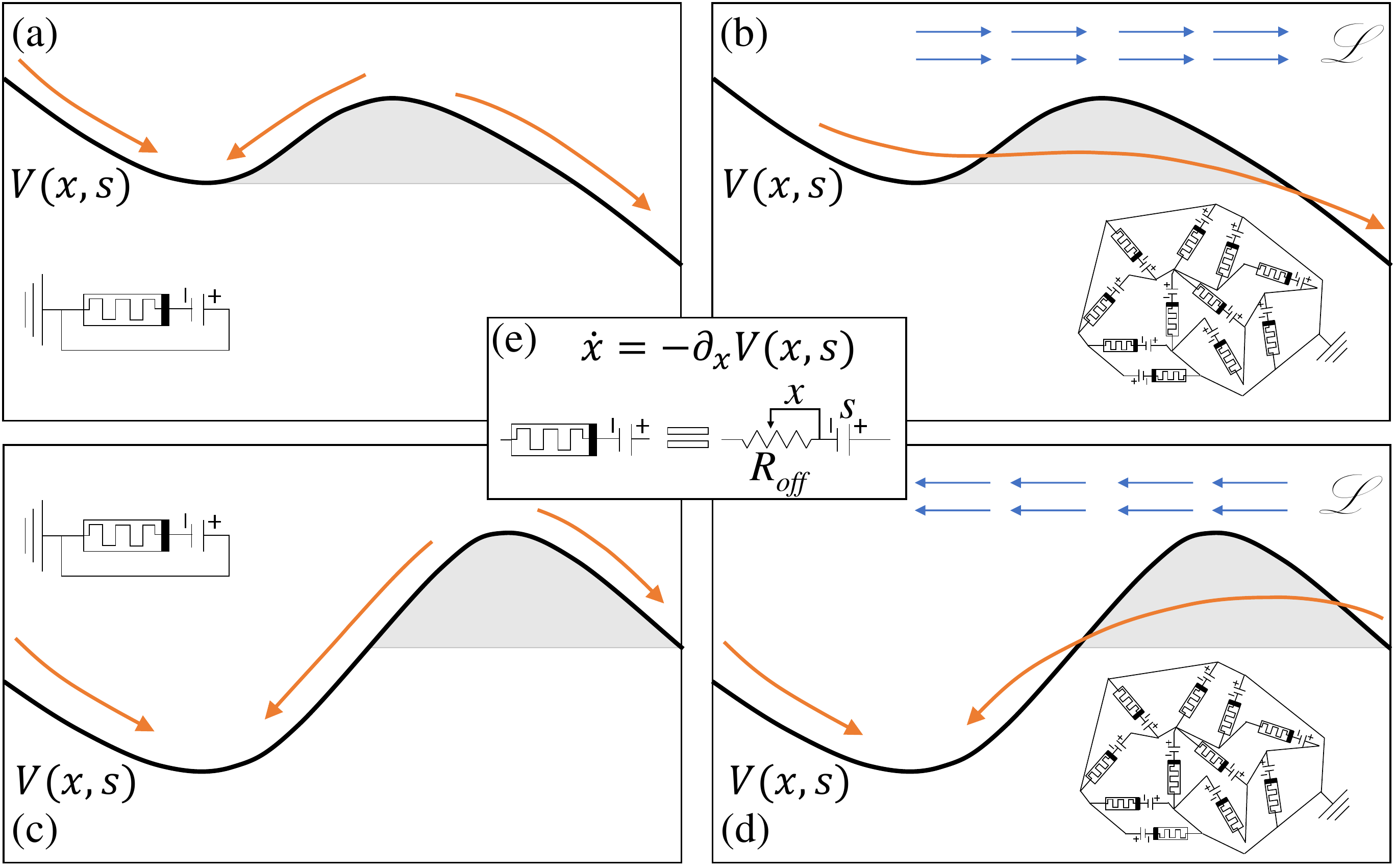}
    %
    \caption{Representation of the tunneling phenomenon. On the left (a,c), the memristor potential $V(x,s)$ for an high (a) and a low (c) value of the voltage $s$ is sketched and two attractors are highlighted through the directions that internal memristor state take depending on the initial conditions. On the right (b,d), the memristor potential $V(x,s)$ now represents either each one of memristors in the network in absence of the coupling, or equivalently the mean field approximation of the system. In this case, the emergent Lyapunv force (blue arrows) pushes the system towards the global minimum creating an effective tunneling through the barrier. In the middle inset (e) the equivalent schematic of the memristive system is reported highlighting the role of the potential $V(x,s)$.  
     }
    \label{fig:figurepot}
\end{figure}

Tunneling phenomena are of current interest for their application to computation, as in quantum annealing~\cite{Oliver2019}. Metastability and tunneling  are a leitmotif for optimization schemes both in a classical and a quantum framework
\cite{qannealer,Hamerly_2019,tunnelingbald}. In addition to quantum annealing, 
interest in alternative approaches to computation and optimization is mounting \cite{Hennessy2019,Vadlamani2020,traversa,Sutton2017,Torrejon2017,toroc2,Isingmachine,Pierangeli_2019}. For instance, there are proposals to use oscillators or more general encodings to process information in the frequency domain~\cite{Isingmachine,Pierangeli_2019,Vadlamani2020,Csaba2020}. Others leverage memory, ranging from near-memory \cite{Singh2019} to in-memory computing, \cite{Ielmini2018,DCRAM,Sebastian2020} and finishing in memcomputing \cite{traversa,Ventra2018}. 
These alternatives are trying to more efficiently solve hard problems across optimization \cite{Kirkpatrick83,glover,Dorigo2004,Yang2010} and  machine learning \cite{LeCun_2015}.

\begin{figure}
    \includegraphics[width = \columnwidth]{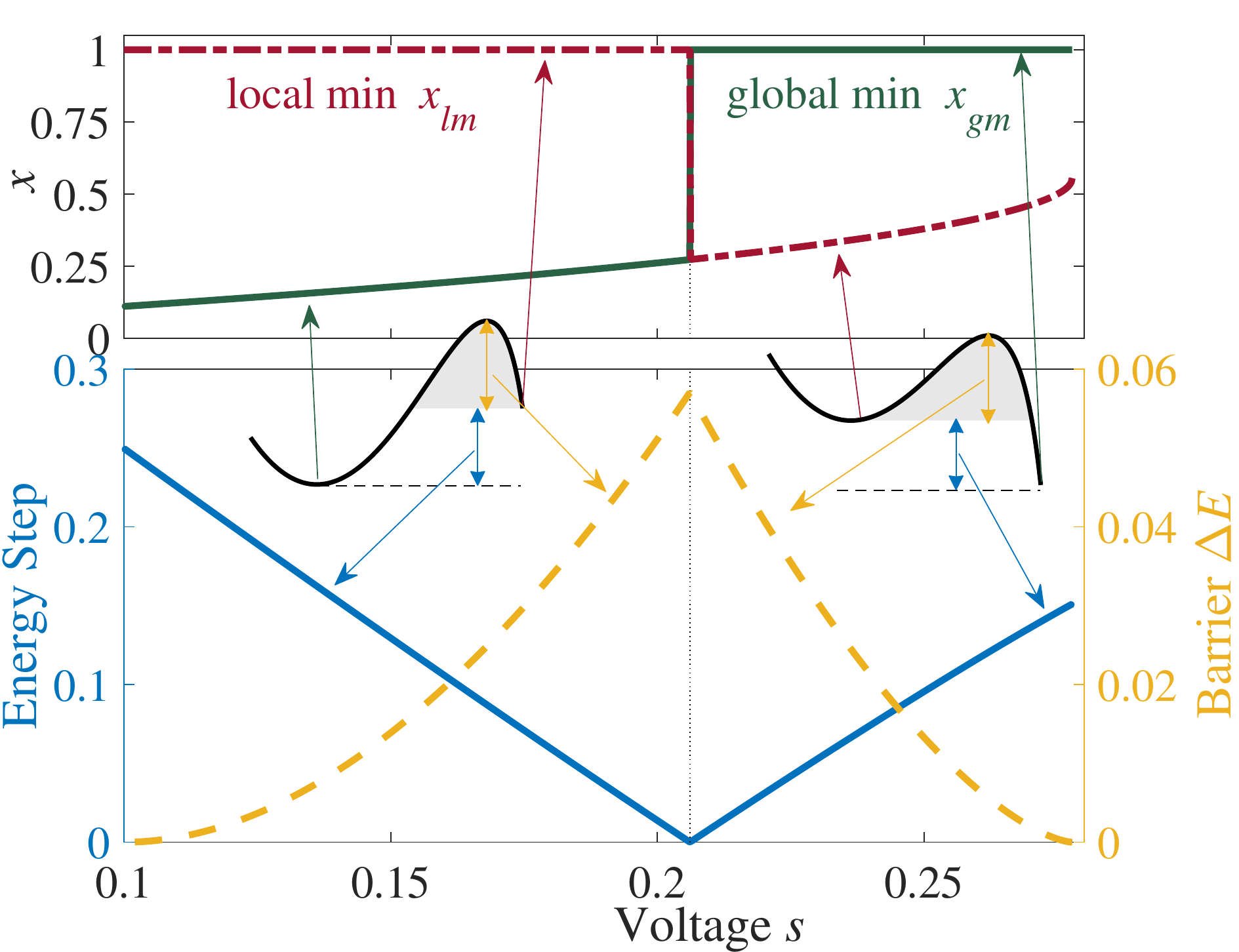}
    \caption{Upper panel: Local ($x_{lm}$, dashed dotted red line) and global ($x_{gm}$, solid green line) minimum location as function of voltage $s$. Lower panel: Energy Step (blue solid curve) and Barrier $\Delta E$ (dashed yellow curve) as a function of voltage $s$.}
    \label{fig:stepbarrier_vs_s}
\end{figure}

\begin{figure*}[ht!]
    \centering
    \includegraphics[scale=0.26]{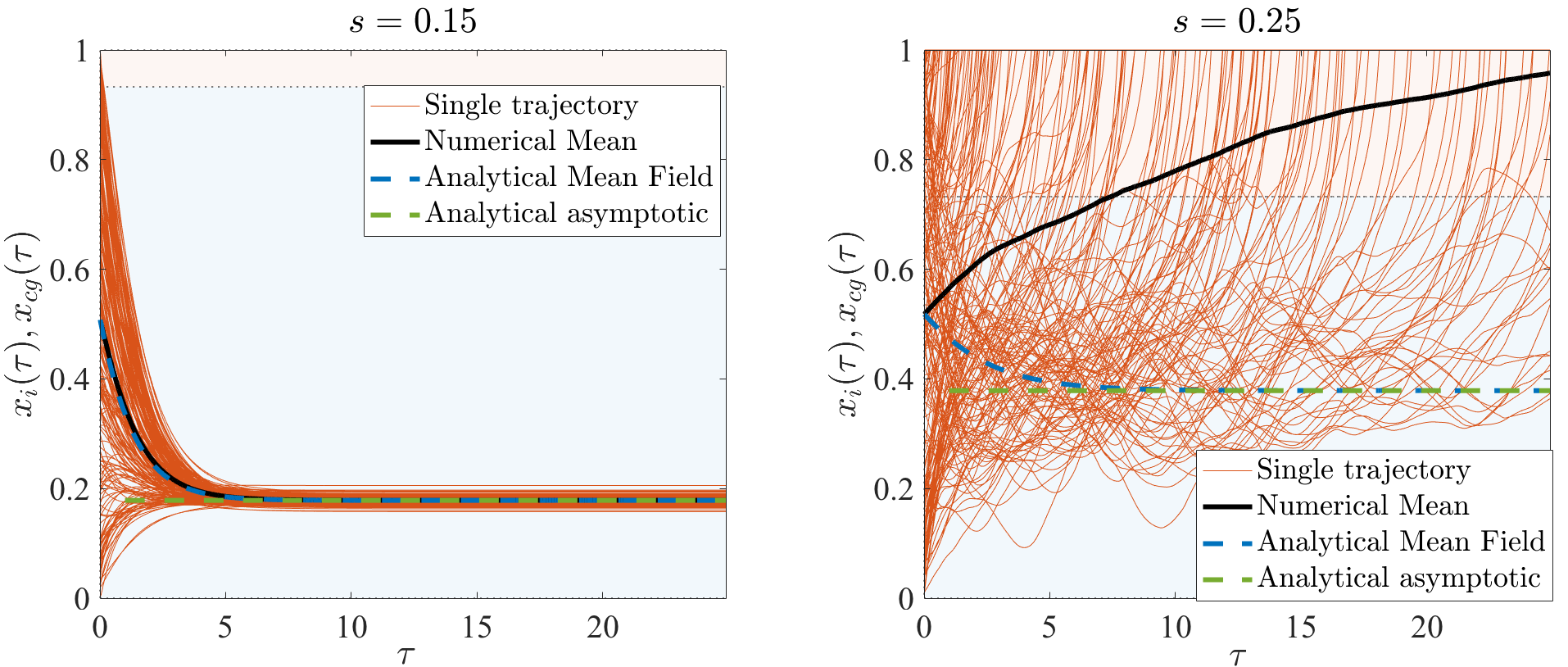}
    \caption{ Dynamics of the system for $s=0.15$ and $s=0.25$, and the mean field basins of the attractions of the effective potential  are represented by the background blue and red coloring. The initial condition for the elements of $x$ is a uniform distribution in $[0,1]$. \textit{Left panel:} trajectories of the dynamical system in the case with $s=0.15$ and $\chi=0.9$ and in line with the mean field regime (equivalent to a single memristor).  \textit{ Right panel:} rumbling trajectories of the dynamical system for the case of the non-convex potential, for $s=0.25$, and in which the mean field equations fail to capture the dynamics of the network. }
    \label{fig:dynamics}
\end{figure*}

Among these alternatives, interest in active or chaotic systems has experienced a revival, also because of their applicability to optimization problems \cite{toroc0}. Chaotic dynamics have been posed as an obstacle to reaching asymptotic fixed points in dynamical system-based computing \cite{tchaos} but there has also been some indirect evidence that unstable dynamics can improve the efficiency of some optimization schemes. In particular it has been noted  that in crossbar arrays made of memristive elements, unstable dynamics can lead to escaping local minima \cite{yang2020} in memristor-implemented simulated annealing.


While escape phenomena are familiar in quantum and thermal settings, there are no examples, of which we are aware, of classical dynamical systems which exhibit barrier escapes when the system is a-thermal \textit{and} passive.
In this article we present an example of barrier tunneling as an emergent, multi-particle effect in an a-thermal, passive \cite{grivet2015passive} system moving in an effective potential.  Both thermal and quantum tunneling involve a distribution which is either intrinsic (quantum) or induced by uncertainty (thermal). Here this distribution is replaced by a coarse-grained representation of the system, which transitions from a laminar phase (always negative Lyapunov exponents) to an unstable or transiently unstable regime (local Lyapunov exponents positive for short transients).  

\textit{The model.}
The example considered is inspired by the dynamics of memristor networks \cite{chua71,chua76,stru08,reviewCarCar,diventra13a,reservoirmem}, in which a low dimensional effective potential representation shows barrier escapes. This property is represented in Fig. \ref{fig:figurepot} and is a novel finding of this paper which we characterize both analytically and numerically. 
This effect comes from the intrinsic coupling among elements, which in the memristor network interpretation is induced by Kirchhoff's laws. Due to the non-linear nature of the coupling, we use a recently obtained exact large-$N$ formula for the matrix inverse, and show analytically that the effective coarse grained dynamics can be approximated by the one of a single memristor (e.g. a mean field approximation). When the effective potential lacks convexity, the emergent Lyapunov force pushes the system into the absolute minimum of the potential, thus effectively tunneling through the  barriers. Interestingly, this Lyapunov force is directly related to instabilities in the dynamics.

The dynamical equation for the circuit has been derived in \cite{Caravelli2017b}, but also applies to physarum molds \cite{johannson2012slime} or as a dynamic generalization of supply-chain or input-output models.  In particular, it models a flow network which obeys current and energy conservation and whose edge dynamics are linear in the currents.
A resistor with memory can be described by an effective dynamical resistance depending on an internal parameter $x$. For instance, $TiO_2$ memristors are approximately described by the functional form $R(x)=\Roff (1-x)+x \Ron$, where $\Ron<\Roff$ are the limiting resistances, and $x\in[0,1]$. The internal memory parameter evolves according to a simple equation of the form $\frac{d}{dt}x=\frac{R_{off}}{\beta} I-\alpha x=\frac{R_{off}}{\beta} \frac{V}{R(x)}-\alpha x$. The parameters $\alpha$ and $\beta$ are the decay constant and the effective activation voltage per unit of time respectively, and determine the timescales of the system. For a single memristor under an applied voltage $S$ we use Ohm's law $S=R I$ to obtain an equation for $x(t)$ in adimensional units ($\tau=\alpha t$) given by
\begin{eqnarray}
 \frac{d}{d\tau}x=\frac{S}{\alpha \beta} \frac{1}{1-\chi x}-x=-\partial_x V(x,s).
 \label{eq:oned}
\end{eqnarray}
Here we have defined $\chi=\frac{\Roff-\Ron}{\Roff}$, with $0\leq \chi \leq 1$ physically, and $V(x, s)$ as an effective potential.

The dynamics of the one-dimensional system of Eqn. (\ref{eq:oned}) are fully characterized by gradient descent in the potential 
\begin{equation}
V(x,s)=\frac{1}{2} x^2 +\frac{s}{\chi} \log(1-\chi x), 
\label{eq:potential}
\end{equation}
with $s=\frac{S}{\alpha \beta}$. The potential can have two local minima and the energy step between the two is shown in Fig. \ref{fig:stepbarrier_vs_s} for various values of $s$ and for $\chi=0.9$. 
We consider $s$ on the restricted interval for which a barrier exists as shown in Fig.~\ref{fig:stepbarrier_vs_s}. In this range, and with $\chi$ near 1, the local minimum can move inside the domain $[0,1]$, and an unstable fixed point (i.e. the location of the peak of the barrier) emerges, leading to two basins of attraction (Fig.~\ref{fig:stepbarrier_vs_s}). For $\alpha=\beta=1$ and $\chi=0.9$ this range is $1/10< s <5/18$ . The requirement that $\chi$, which characterizes the nonlinearity of the system, be near 1 implies that the phenomenon is nonperturbative. The asymptotic behavior of this single-element dynamical system is fully characterized by the simple basins of attraction of the potential, and presents no exotic features \footnote{ Typically, since one must have $0\leq x\leq 1$, the equations of motion of the single variables are supplied with (non-absorbing) cutoff functions, e.g. $\frac{d}{d\tau} x=-W(x) f(x)$, with $W(x)=1$ for $0\leq x\leq 1$ and zero otherwise. }. An analytical solution of such differential equation has been recently obtained in implicit form \cite{coffrin}.

When, instead of a single memory device, we have a circuit composed of many memristors, each with resistance $R(x_i)$ and with voltage generators $S_i$ in series, (taken to be constant $S_i=\tilde S$),
the differential equation (\ref{eq:oned}) generalizes to a system of coupled and nonlinear ordinary differential equations.
The network dynamics equation for the memory elements $x_i(t)$ is \cite{Caravelli2016rl,Caravelli2019Ent}:
\begin{eqnarray}
\frac{d}{dt} \vec x=\frac{1}{\beta}(I-\chi \Omega X)^{-1} \Omega \vec S-\alpha \vec x,
\label{eq:manyd}
\end{eqnarray}
with $\chi=\frac{\Roff-\Ron}{\Roff}< 1$, and $X_{ij}(t)=x_i(t) \delta_{ij}$. The matrix $\Omega$ is the projection operator ($\Omega^2=\Omega$) on the vector space of the cycles  of $\mathcal G$, the graph representing the circuit~\cite{Caravelli2016rl}. In practice, $\Omega$ can be determined via the directed incidence matrix $B$ of $\mathcal G$ as $\Omega=I-B(B^t B)^{-1} B^t$ \cite{Caravelli2016ml,Caravelli2017b,Zegarac,Nilsson,bollobas2012graph}. That $\Omega$ is a projection operator is a mathematical representation of Kirchhoff's circuit laws. In this paper we choose $B$ to be a random matrix in order to abstract the dynamical system from a particular circuit topology.

The system has a Lyapunov function when $\vec S$ is constant (see Supp. Mat. A), and memristors are passive elements and thus cannot possess positive Lyapunov exponents (e.g. the system cannot be unstable for long times \cite{grivet2015passive}). Passive components subject to DC voltage generators will approach an equilibrium and thus only transient forms of instability are possible. Exact information about the behavior of Eqn.~(\ref{eq:manyd}) has been difficult to obtain as the matrix inverse $(I-\chi \Omega X)^{-1}$ contains the variables $x_i(t)$ and it is therefore difficult to solve analytically. Additionally,  B\'{e}zout's theorem for quadrics of order $2$ with $N$ variables suggests the number of equilibrium points (stable, unstable or saddle) can be exponential in the number of memristors \cite{coffrin} (this can be justified intuitively, as each memristor has 2 fixed points).
Despite this complexity, the higher-dimensional dynamics of \textit{networks} of memristors can be well-approximated as an effective one-dimensional system in a coarse-grained representation as we show in the following section.

Interestingly, in the regime where the potential $V(x,s)$ has multiple minima and the initial condition of the system lies outside of the attraction basin of the global minimum, the system exhibits instability. This can be seen in Fig. \ref{fig:dynamics} (red curves), where for $\chi=0.9$ the system shows qualitatively different dynamics at different values of $s$. Here we have assigned the initial conditions of the elements of $\vec{x}$ according to a uniform distribution on $[0,1]$ and considered two cases for $s$: $s=0.15$ for which the global minimum is located at $x_i\approx 0.2$, and $s=0.25$ for which the global minimum at $x_i=1$ (see Fig.~\ref{fig:stepbarrier_vs_s}). The former case shows laminar behaviour with the trajectories converging smoothly to $x_i \approx 0.2$. On the other hand, for $s=0.25$, initial conditions for the elements of $\vec x$ are almost all within the ``local" minimum's basin of attraction since now the global minimum is at $x_i=1$; now the system shows unstable dynamics, pushing trajectories to the other side of the barrier and converging to the global minimum $x_i=1$ (see Fig.~\ref{fig:dynamics}).

\begin{figure}
    \centering
    \includegraphics[scale=0.16]{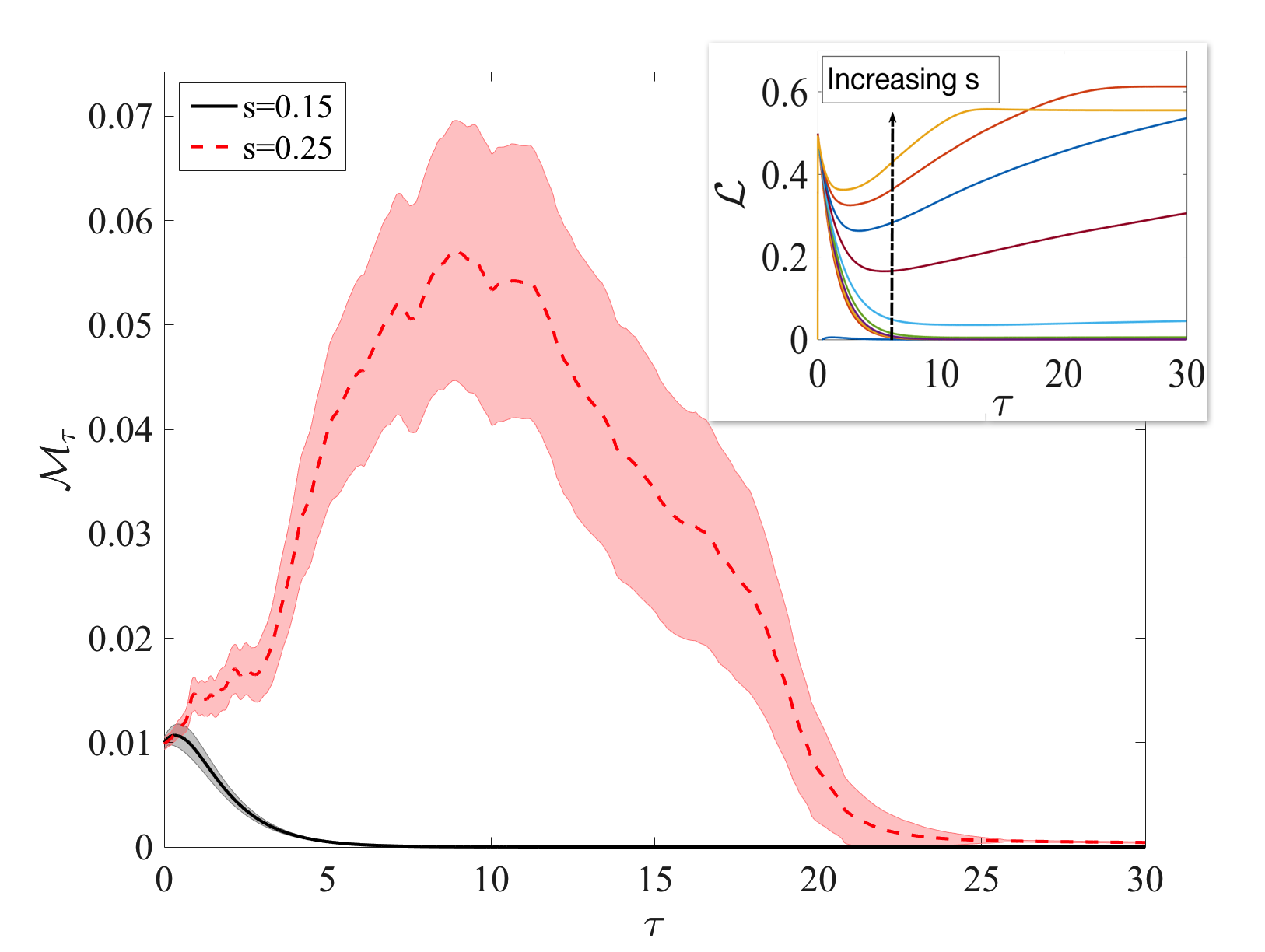}
    \caption{Growth and decay of $ \mathcal M_\tau=\frac{1}{N}\sum_{i=1}^N |x_i(t)-x^{\epsilon}_i(t)|$ as a function of time for $\chi=0.9$ and $s=0.15$ and $s=0.25$. The realization of the noise $\epsilon$  on the initial condition is such that initially $\mathcal M_0=0.01$. In the laminar regime $(s=0.15)$, $\mathcal M_\tau$ quickly decays to zero, while it grows for a transient of time in the unstable regime.   In order to measure the growth and decay of $M_\tau$, we look at $\lambda=\frac{1}{\bar \tau} \log \frac{\mathcal M_{\bar \tau}}{\mathcal M_0}$. For $s=0.15$, we obtain $\lambda= -0.79(1)$ measures on the whole interval, while for $s=0.25$ we obtain $\lambda=0.2(1)$, and thus positive. \textit{Inset:} Evolution of the Lyapunov force as a function of time and for $s\in[0.15,0.27]$. Every curve is averaged over $100$ initial conditions and the shadow area represents the standard deviation on the mean curves.}
    \label{fig:mt}
\end{figure}

We quantify this behavior by examining the distance between copies of the system with similar initial conditions $\vec x^{(1)}(\tau)$ and $\vec x^{(2)}(\tau)$. As a distance, we use the absolute value, which is essentially an $L_1$ measure of the Lyapunov exponents for the dynamics. For these experiments, we  initialize the two systems in the same state $\vec x_0$, but with a small deviation $\vec \epsilon$, with $||\vec \epsilon||_{1}=0.01$, and where $||\cdot ||_{1}$ represents the 1-norm. It is natural to define the quantity $\mathcal M_\tau=\frac{1}{N} \sum_{i=1}^N |x_i^{(1)}(\tau)-x_i^{(2)}(\tau)|$, e.g. a normalized  $1$-norm which quantifies the deviation of the trajectories element by element, and which is convenient because it is naturally normalized between $0$ and $1$. The results are shown in Fig. \ref{fig:mt}, where we see that for the case in which $s=0.15$ (laminar), $\mathcal M_\tau$ decays to zero rapidly. In the case $s=0.25$ (unstable), $\mathcal M_\tau$ grows for a transient of time of approximately 7-fold longer before asymptotically reaching zero.  For this reason, we dub this phenomenon as a \textit{rumbling transition} due to the chaoticity of the individual trajectories, resulting from the nonlinearity of the interaction.

\textit{Coarse-grained dynamics.} We now make the intuitive picture above more precise by introducing an effective mean field potential derived from an approximation for the matrix inverse. It has been recently proven that the resolvent  of large matrices exhibits statistical regularities, for which it can be approximated by an effectively one-dimensional matrix \cite{BCCV} which is universal at the zeroth-order in the limit of weak correlations. The result is rather general and has applications to various domains of complexity science \cite{BCCV2}.
If we define $\vec f=\Omega \vec x$ and the matrix
\begin{eqnarray}
\tilde A=\begin{pmatrix}
        f_1(\vec x) & \cdots & f_1(\vec x)\\
        f_2(\vec x) & \cdots & f_2(\vec x)\\
        \vdots & \ddots & \vdots\\
        f_N(\vec x) & \cdots & f_N(\vec x)
        \end{pmatrix}= \vec f\otimes \vec 1^t,
\end{eqnarray}
we then have the approximate relation  (details in Supp. Mat. B)
\begin{eqnarray}
(I-\chi\Omega X)^{-1}= I+\frac{1}{N} \frac{\chi}{1-\frac{1}{N}\chi\sum_{i=1}^N f_i(\vec x)}\tilde A +O\left(\frac{1}{N}\right).
\label{eq:meanfieldexp}
\end{eqnarray}

Thus, if we define a coarse-grained variable $x_{cg}=\frac{1}{N} \sum_i f_i$, and  define $\langle \Omega \vec S\rangle=\frac{1}{N} \sum_{i=1}^N (\Omega \vec S)_i$, we can write an effective one dimensional dynamics:
\begin{eqnarray}
\frac{d}{d\tau} x_{cg}&=&\frac{1}{\alpha \beta}\Big( \langle \Omega \vec S\rangle+\frac{\chi \langle \Omega \vec S \rangle}{1-\chi x_{cg}}x_{cg}\Big)- x_{cg}+\mathcal L(\vec x) \nonumber \\
&=&-\partial_{x_{cg}} V(x_{cg},\chi) +\mathcal L(\vec x)
\label{eq:meanfielddynp}
\end{eqnarray}
where $\mathcal L(\vec x)$ is an effective force due to the fact that the coarse-graining is not exact. We thus see that at the zeroth order the mean field variable obeys dynamics similar to those of a single memristor, where now the parameter $\frac{s}{\alpha \beta}$ is replaced by the mean field value $s=\frac{\langle \Omega \vec S\rangle}{\alpha \beta}$.
Thus, the equation above again represents gradient descent dynamics for a one-dimensional system, but where  an effective external force $\mathcal L(\vec x)$ emerges from the interaction between the large set of (hidden) variables $x_i$, and which we can evaluate numerically on the dynamics. In a sense, while the $N$ particles somehow feel a similar potential, these can interact as well, leading to a non-trivial collective dynamics. As a result we find that the variable $x_{cg}$ is a natural dynamical order parameter to study for our system.

\textit{Laminar versus Transient unstable dynamics.}  We numerically examined the validity of the coarse-grained dynamics for various ranges of the parameters $s$ and $\chi$.\footnote{The numerical results are obtained using a simple Euler integration with a step $dt=0.1$, but the results are robust to smaller values, and agree with independent simulations using Runge-Kutta 4 method and similar step sizes.} 

The picture which emerges is represented in Fig. \ref{fig:dynamics}. As discussed previously, we consider a uniform distribution in [0,1] for the initial condition of the elements of $\vec x$ and the laminar case $s=0.15$ for which the global minimum is located at around $x_i\approx 0.2$ and the unstable case $s=0.25$ with the global at $x_i=1$ (see Fig.~\ref{fig:stepbarrier_vs_s}). In the former case the majority of the initial conditions, $x_i(0)$ lie in the basin of the global minimum. On the other hand, for $s=0.25$, almost all initial conditions for the elements of $\vec x$ are on the basin of the ``local" minimum as now the global minimum is at $x_i=1$. In the laminar regime the mean-field dynamics accurately fit the dynamics of the system. This is always the case for $\chi\ll 1$, in which very few trajectories cross the barrier and the dynamics are therefore smooth.  In this regime, the effective force $\mathcal L(x)$  (which we call Lyapunov force for reasons which become clear soon) is typically small after a short transient, as we can see the inset in Fig. \ref{fig:mt}.

This implies that the equilibrium points of the dynamics, which are those satisfying $\frac{d}{d\tau} x_{cg}=0$, are well approximated by the mean field dynamics (e.g. the particles follow the same basin of attraction). This gives the equilibrium value
\begin{eqnarray}
x_{cg}^*=\frac{1}{N}\sum_{ij} \Omega_{ij} x_j^*=\frac{\alpha \beta-\sqrt{\alpha^2\beta^2-4 \alpha \beta \chi \langle \Omega \vec S\rangle}}{2 \alpha \beta \chi},
\end{eqnarray}
which matches simulations.
The equation above is equivalent to finding values of $x_{i}\in [0,1]$ such that $\vec x\cdot \vec c=a$, and $\vec x$ is such that $\vec x=\Omega \vec x$ (See Supp. Mat. C). This equation has many multiple possible solutions in high dimensions, explaining the large number of asymptotic states of the memristive dynamics commonly seen in numerical experiments \cite{caravellisheldon}. Yet, this also implies that the dynamics can be succinctly described by a scalar order parameter for the system, and that the mean field and the effective potential do play a role. 

\begin{figure}
    \centering
    \includegraphics[scale=0.17]{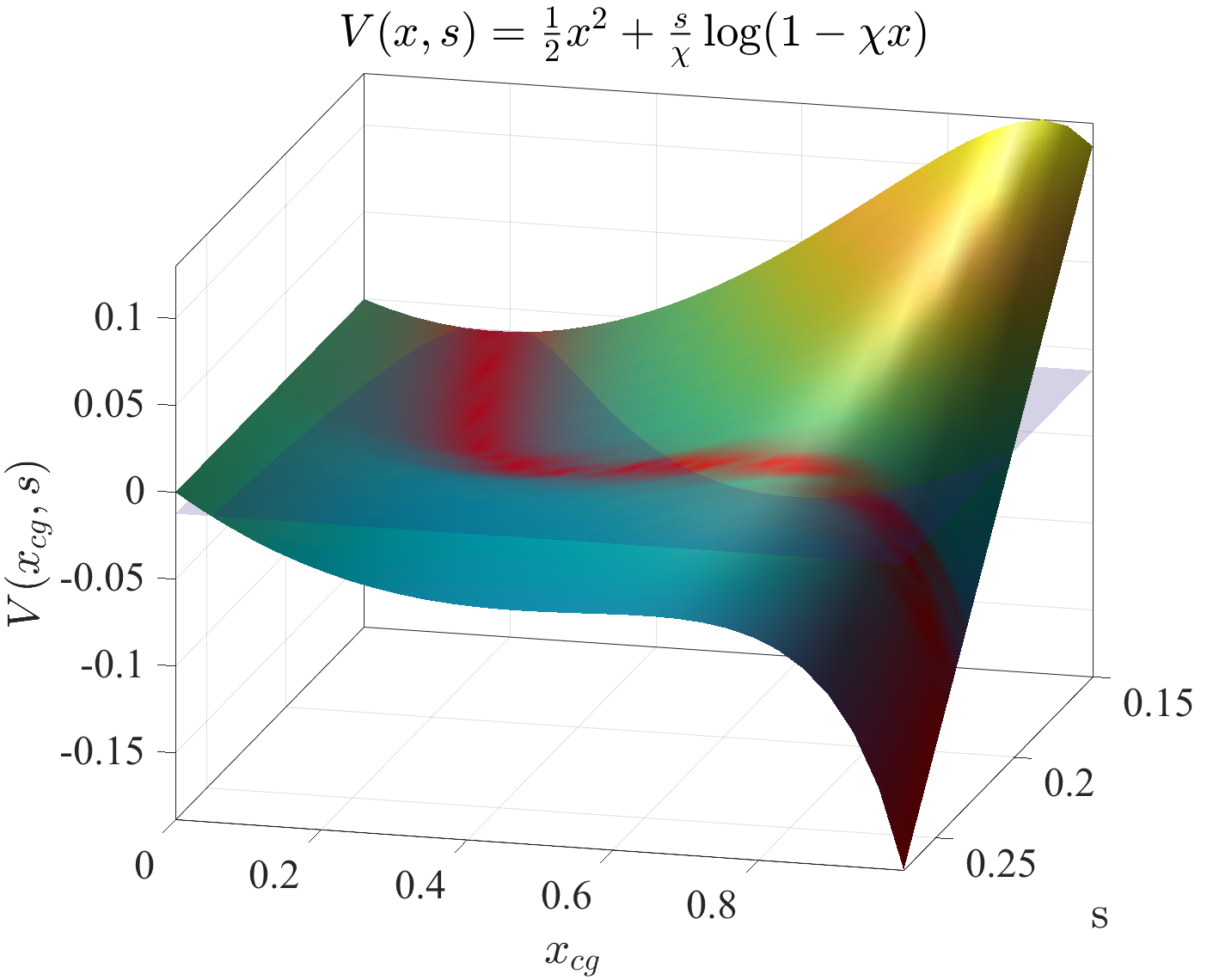}
    \caption{Average asymptotic state $E\left[ \langle \vec x(t=\infty)\rangle \right]$ (red curve) of the memristor coarse grained variable obtained from 400 Monte Carlo instances for each point $(s,\vec x)$, on a grid of $50$ points $s\in[0.15,0.27]$ and 30 points $x_i\in[0,1]$ and for $\chi=0.9$, obtained with an Euler integration method with $dt=0.1$, and tested against a Runge-Kutta 4 integration method for stability. The transparent plane is for visual aid to show that the average climbs the barrier.}
    \label{fig:instanton}
\end{figure}


\begin{figure*}
    \centering
    	\begin{overpic}[height=9.5cm]{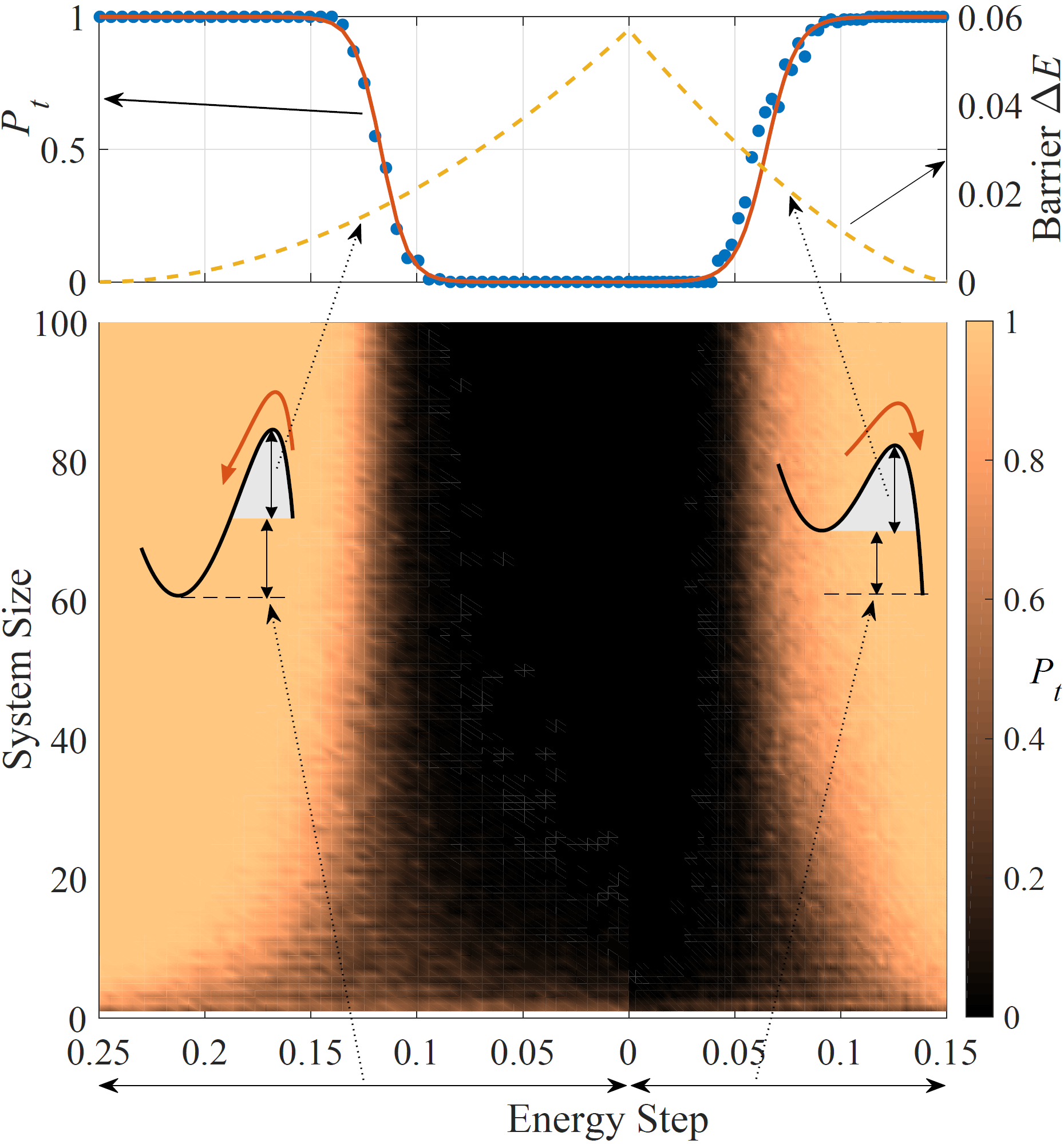}
		\put (0,98) {(a)}
		\put (0,65) {(b)}
	\end{overpic}
	\hspace{.4cm}
    \begin{overpic}[height=9.5cm]{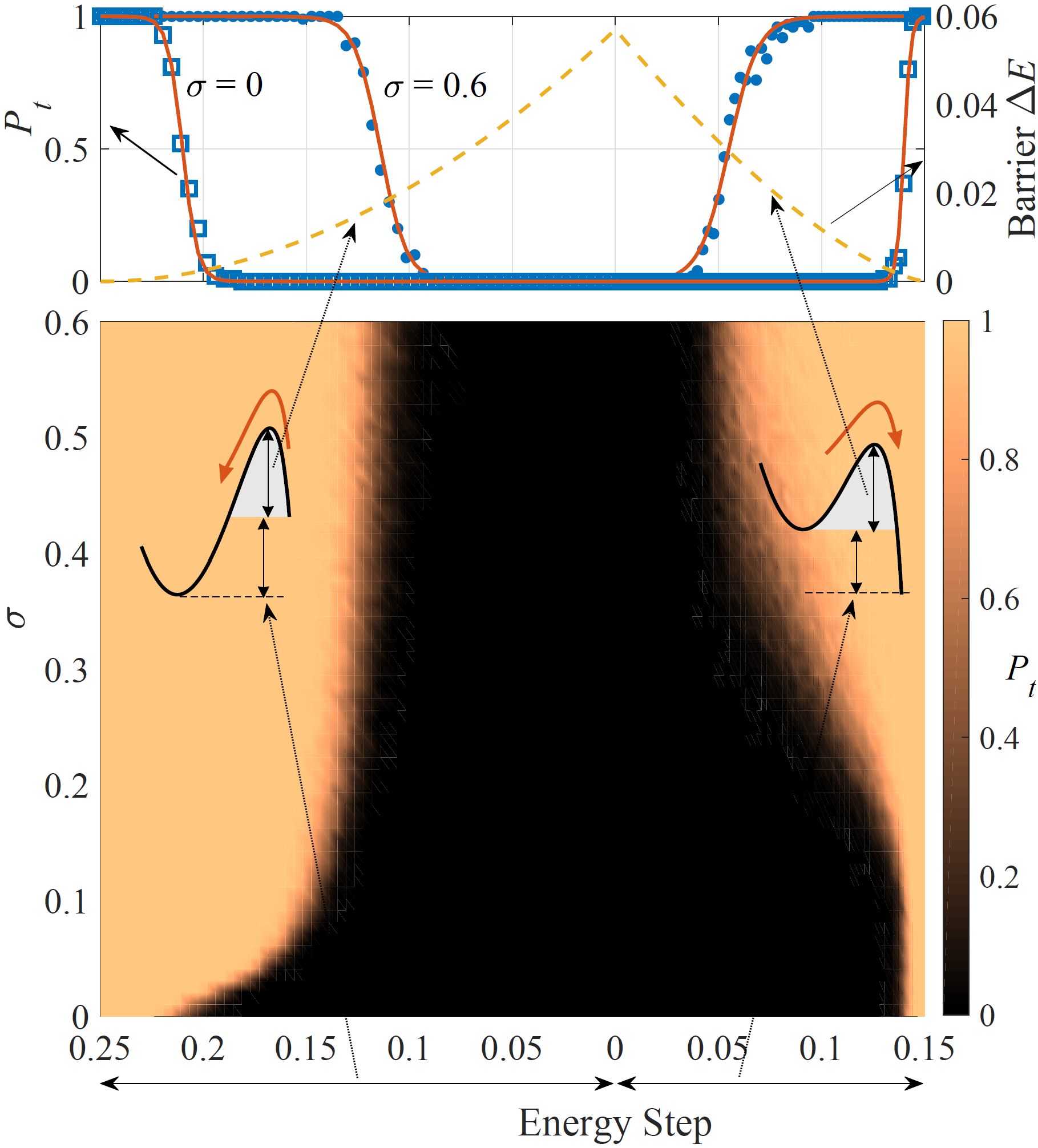}
    	\put (0,98) {(c)}
		\put (0,65) {(d)}
	\end{overpic}
	\caption{Tunneling probability $P_t$ from the local to the global minimum of $V(x,s)$. Insets (a-b) are evaluated for the standard deviation of initial condition distribution around the local minimum $\sigma=0.5$. (b) probability $P_t$ as a function of the energy step between the two minima and the size of the system (or equivalently the number of memristors). (a) section of the plot in (b) for System Size = 100. Blu dots are the numerical evaluation of $P_t$ vs the energy step while the solid red curve is the fitting using the $\tanh$ functions. The dashed yellow curve is the barrier $\Delta E$ as a function of the energy step. Insets (c-d) are evaluated for System Size = 100. (c) probability $P_t$ as a function of the energy step between the two minima and the standard deviation $\sigma$ of initial condition distribution around the local minimum. (c) section of the plot in (d) for $\sigma= 0$ and 0.5. Blu dots and squares are the numerical evaluation of $P_t$ vs the energy step while the solid red curves are the fittings using the $\tanh$ functions. The dashed yellow curve is the barrier $\Delta E$ as a function of the energy step.}
    \label{fig:tunnellingPt}
\end{figure*}

The ability of the coarse-grained dynamics to capture the memristor network evolution changes abruptly when most $x_i$ have initial conditions outside of the attraction basin of the global minimum, as can be seen in the right panel of Fig. \ref{fig:dynamics}. Fixing $\chi=0.9$ and  $s= 0.25$ such that the global minimum is now at $x_i=1$ the one-dimensional representation which we obtained at the leading order fails to captures the dynamics of the system. Individual trajectories $x_i(t)$ now deviates from the coarse-grained value $x_{cg}(t)$ substantially, as we can see in Fig. \ref{fig:dynamics} (right), with clear instabilities which lead them to the absorbing boundary $x_i(t)=1$.  While this might seem irksome at first, we  argue that this phenomenon is worthy of attention. In fact, the effective Lyapunov force $\mathcal L(x)$ is now consistently large along the dynamics, pushing the system towards the absolute minimum. Since the resolvent expansion of Eqn. (\ref{eq:meanfieldexp}) is valid in the limit of weak correlations of the matrix elements, the fact that the Lyapunov force is non-zero is naturally associated to strong correlations between the dynamical variables.

As described, the system displays this behaviour even when randomly initialized, and was investigated systematically. To highlight the role of the basin of attraction of the global minimum we employed Monte Carlo simulations, randomizing over the initial conditions on $[0,1]$. The results are shown in Fig.~\ref{fig:instanton}. The surface shows the potential $V(x,s)$ as a function of $x_{cg}$ and $s$. The superimposed red curve represents the asymptotic average position $E\left[ \langle \vec x(t=\infty)\rangle \right]$ where $E\left[\cdot\right]$ is the average over the Monte Carlo samplings and $\langle \vec x\rangle =\frac{1}{N}\sum_ix_i$ is the average over the components. As can be seen, before the barrier disappears the dynamics reaches the right minimum of the potential $V(x,s)$ with probability one.  
A glimpse of the structure of the basins of attraction can be obtained by fixing the initial conditions homogeneously for all but two variables $x_i$, and look at the resulting asymptotic state in these two variables (see the Supp. Mat. E for details) near the transition point.

\textit{Tunneling.} To capture this as a tunneling effect, we initialized the system around the local minimum using the initial condition $\vec x(0)=x_{lm}(s)+\vec \sigma$ where $x_{lm}(s)$ is the location of the coarse grained local minimum as depicted in figure \ref{fig:stepbarrier_vs_s}, and $\vec \sigma$ being a random vector drawn from the Gau\ss ian distribution $\mathcal N(0,\sigma)$. Given this, we performed Monte Carlo simulations and obtained the probability $P_t$ that the system hops from the local minimum to the global minimum, or, in other words, the probability $P_t$ that the initial condition $\vec x(0)=x_{lm}(s)+\vec \sigma$ leads to $\vec x(t=\infty) = x_{gm}(s)$. We also note that $x_{lm}(s)$ shows an abrupt change for $s=s_{crit}\approx 0.206$ as depicted int Fig.~\ref{fig:stepbarrier_vs_s} as at that point the local and global minima switch locations. This leads to two different effective tunneling directions for $s>s_{crit}$ and $s<s_{crit}$ (see the sketch in Fig.~\ref{fig:figurepot}). 

We first investigated how the tunneling emerges as a collective behaviour of the system. To this end we evaluated the dynamics of systems as the number of elements increased. Fig.~\ref{fig:tunnellingPt}-b reports the tunneling probability for $\sigma=0.5$ and for system sizes from 1 to 100 components and for the entire range of $s$. We can observe that at small system size the tunneling probability is proportional to the number of initial conditions that fall into the global minimum attraction basin to end with an almost perfect sigmoid ($\tanh$), as a function of the energy barrier $\Delta E$ (see also Fig.~\ref{fig:tunnellingPt}-b). This transition is smooth and qualitatively equivalent for both tunneling directions. Therefore the system shows a size-depended transition from gradient dynamics to a collective tunneling towards the global optimum, led by the emergent Lyapunov force. 

We also investigated how the emergent force depends on the spread of the initial conditions around $x_{lm}$, i.e., how it depends on $\sigma$. Figure \ref{fig:tunnellingPt}-d reports the tunneling probability $P_t$ as a function of $\sigma$ for a system of 50 components. An interesting feature manifests itself: for $\sigma=0$ there is a non-null height of the barrier $\Delta E$ for each tunneling direction below which the system is still able to tunnel towards the global minimum with probability 1 (this is highlighted in figure  \ref{fig:tunnellingPt}-c). This shows that the Lyapunov force is an intrinsic collective large-system-size feature present even in the absence of any randomness in the initial conditions. 

Based on the latter finding, we investigated whether escape from the minimum might be due to a transient instability of non-normal dynamics \cite{TrefethenEmbreeBook2005}. One possible mechanism would in fact be that via this (local) transient instability perturbations are amplified for a short time and the system reaches the basin of attraction of another minimum. To examine this we studied the local stability of the system near $x_{lm}$ via the Jacobian matrix. While the details are provided in App. D, we analyzed a large number of local minima via Monte Carlo, finding that the non-normality of the Jacobian does not lead to any amplification phenomenon. This also leaves us with the only option that the instability of Fig. \ref{fig:mt} is in fact due to a non-perturbative and cooperative phenomenon between the dynamical variables. From the point of view of the effective potential $V(x,s)$, this analysis implies that the global minimum lying behind a barrier is reached (via the instabilities above mentioned) via a dynamics assisted by the effective Lyapunov force which pushes the system towards it. 


\textit{Conclusions.} 
Traditional escape phenomena in thermal and quantum systems are a consequence of either thermal fluctuations or the spreading of a wavefunction.  We have presented an additional setting in which barrier escapes emerge in the effective description of a multiparticle system. We presented a class of dynamical system, derived from networks of memristors, that can be mapped to an effective one-dimensional system. The resulting dynamics are characterized by a potential which depends on the (mean-field) external applied voltage, and an effective force which occurs when the system becomes Lyapunov unstable. The result of this instability is that the effective representation may jump between basins of attraction, even through a barrier. The explanation we provide is simple but difficult to capture analytically and is due to the large number of directions in which the system can travel, via a sequence of saddle points in the dynamics, and from which the local instability emerges. 
This is compatible with previous investigations of memristive circuit dynamics applied to logic gates
\cite{DMM2,Traversa2015}, and the observation of instanton-like behavior in the dynamics \cite{Inst1,Inst2}.

Our analysis of the effective ``Lyapunov" force provides an explanation of the observed tunneling as an epiphenomenon of the interaction between the memory elements. In principle this might fit into a coarse graining argument of the hidden variables, due to a Zwanzig expansion \cite{zwanzig}, which will be considered in the future. However, we wish to stress that the system we studied is \textit{a-thermal} (\emph{i.e.} no stochastic forces were introduced in our analysis) and that the effective Lyapunov force is far from random. 

The main message of this paper is that the introduction of hidden variables in a dynamical system
can lead to transitions between local and global minima of the effective description via instabilities in the full system. In this sense, this type of behavior is novel and worth further exploration. 
In quantum and statistical field theory, tunneling phenomena are described by instantons, e.g. solutions in which the field tunnels in a finite time between local minima. Instantons, originally introduced in cosmology \cite{Coleman}, are nonlinear solutions to the field equation which are non-local in time, and represent tunneling between two local minima. Typically, instantonic behavior is associated to some form of stochasticity, whether thermal or quantum in nature. In this paper we have shown that instantonic behavior can also be an emergent phenomenon \cite{Anderson72} thanks to the nonlinear interaction between many components.
We hope that our paper can spark further interest in the study of tunneling in dynamical systems via ``hidden variables", with their interesting properties, and with important applications to optimization \cite{yang2020,diventra13a,Inst1,Inst2}, and machine learning \cite{sompolinsky,ganguli}.

\textit{Acknowledgements.}  We acknowledge the support of NNSA for the U.S. DoE at LANL under Contract No. DE-AC52-06NA25396.  FC was also financed via DOE-LDRD grants PRD20170660 and PRD20190195, and would like to thank S. Bartolucci, F. Caccioli and P. Vivo for collaborations on the resolvent formula which enabled us to write this paper, and C. Coffrin and Y.-T. Lin for comments on the draft. FCS was also supported by a Center for Nonlinear Studies fellowship.

\bibliographystyle{ieeetr}
\bibliography{Tunneling}

\clearpage
\ \\
\appendix
\onecolumngrid
\begin{center}
\textbf{\Large Supplementary Material}
\end{center}

\section*{A: Proof of Lyapunov function for sharp boundaries} 
In this section we provide a proof of the fact that the system is dissipative when each state variable is far from the boundary $x_i=1$ \cite{coffrin,caravellisheldon}. We consider the case of infinitely sharp boundaries, e.g. $\lim_{p\rightarrow \infty} W_p(x)=\theta(x)\theta(1-x)$. In order to prove it, 
we attempt to write a Lyapunov function of the form
for the differential equation (when boundaries are not considered)
\begin{eqnarray}
\frac{d}{dt} \vec x=\frac{1}{\beta}(I-\chi \Omega X)^{-1} \Omega \vec S-\alpha \vec x,
\end{eqnarray}
which we can rewrite as
\begin{eqnarray}
\frac{1}{\alpha}(I-\chi \Omega X)\frac{d}{dt} \vec x=\frac{1}{\alpha \beta} \vec{s}- (I-\chi \Omega X)\vec x,
\label{eq:rewritten}
\end{eqnarray}
where we wrote $\Omega \vec S=\vec {s}$.
\begin{equation}
    L =   \frac{1}{3} \vec{x}^T X \vec{x} - \frac{\chi}{4} \vec{x}^T X\Omega X \vec{x}
    -\frac{1}{2\alpha\beta} \vec{x}^T X\vec{s}.
\end{equation}
After taking a time derivative of the Lyapunov function above, we get
\begin{align}
    \frac{dL}{dt} &=  \dot{ \vec{x}}^T\left( X \vec{x} - \chi X\Omega  X \vec{x} - \frac{1}{\alpha\beta} X\vec{s} \right)= \dot{ \vec{x}}^TX\left(  \vec{x} - \chi \Omega  X \vec{x} - \frac{1}{\alpha\beta} \vec{s} \right)\nonumber  \\
    &= - \dot{ \vec{x}}^T\ X\left(\frac{1}{\alpha\beta} \vec s- (I-\chi \Omega X)\vec x \right) \text{   which is Eqn. (\ref{eq:rewritten})}\nonumber\\
     &=-\frac{1}{\alpha} \dot{ \vec{x}}^T X(I - \chi \Omega X)\dot{ \vec{x}}=-\frac{1}{\alpha} \dot{ \vec{x}}^T (X - \chi X\Omega X)\dot{ \vec{x}}\nonumber\\
    &= -\frac{1}{\alpha}\dot{ \vec{x} }^T \sqrt{X}(I - \chi \sqrt{X}\Omega \sqrt{X})\sqrt{X}\dot{ \vec{x}}= -\frac{1}{\alpha}||\sqrt{X}\dot { \vec{x}}||^2_{ (I - \chi \sqrt{X}\Omega \sqrt{X})} 
\end{align}
We obtain that if $I-\chi \sqrt{X} \Omega \sqrt{X}\succ 0$ (and $X\neq 0$), the Lyapunov function has negative derivative always. This occurs if $x_i\in [0,1]$, which is a key factor in the dynamics, it is not hard to see that $\sqrt{X} \Omega \sqrt{X}\prec 1$ (if $x_i\in [0,1]$), and since $\chi < 1$, the Lyapunov property applies also in this case. The existance of a Lyapunov function also implies that there is a global notion of ``energy" beyond the mean-field potential provided in the main text.

\section*{B: Large-N resolvent and mean field equations}
In this section we discuss the theorem used in the main text to derive the mean field potential. Such theorem has been derived with the intention of studying resolvent matrices for generic random variables whose correlations are weak (in a sense defined below).

\textbf{Theorem.} Let $A=(a_{ij})$ be a random $N\times N$ matrix, characterized by the joint probability density function of the entries $P_A(a_{11},\ldots,a_{NN})$. Let $a_{ij}\geq 0$, $\rho(A)<1$, $\beta>1$ and 
\begin{align}
\langle a_{ij}\rangle_A &= z_i/N>0 \label{scaling1}\\
\langle a_{ij}a_{k\ell}\rangle_A-\langle a_{ij}\rangle_A\langle a_{k\ell}\rangle_A &\sim\frac{C_{ij}}{N^\beta}\delta_{ik}\delta_{j\ell}\qquad\text{as }N\to\infty
\label{scaling2}
\end{align} 
for all $i,j$, with $\langle(\cdot)\rangle_A$ defined as 
\begin{equation}
    \langle (\cdot)\rangle_A=\int\mathrm{d}a_{11}\cdots\mathrm{d}a_{NN}P_A(a_{11},\ldots,a_{NN})(\cdot)\ ,\label{langleA}
\end{equation}
and $C_{ij}$ be constants. 
Then we have:
\begin{eqnarray}
(I-A)^{-1}= I+\frac{1}{N} \frac{1}{1-\bar z}\vec z \otimes \vec 1^t +O(\frac{1}{N^\gamma}).
\end{eqnarray}
where $\bar{z}=(1/N)\sum_{i=1}^N z_i$, and $z_i=\sum_{j=1}^N A_{ij}$, with $\gamma>0$. The results above hold irrespective of the precise form of $P_A$. For a proof, see \cite{BCCV}, and an application of the formula above to complex datasets \cite{BCCV2}. While the theorem is stated for a particular form of the correlation between the elements, as a matter of fact it is more general, and it applies to correlation functions which are sufficiently sparse. Nonetheless, for finite $N$ the expansion has some corrections which are important, and that we estimate numerically in the main text.

In our case, the ensemble $A$ is the one of the random projector operators. We will make hereby the connection between the theorem and our setup.
 The formula for the resolvent is exact when the span of the operator $\Omega$ is one dimensional, as it can be promptly seen using the Sherman-Morrison formula.
 
In order to derive a mean-field (or coarse-grained) dynamics for the network of memristors, we now use a recently obtained formula for the resolvent.
Let us now consider the equation:
\begin{equation}
    \frac{d}{dt} \vec x=\frac{1}{\beta}(I-\chi \Omega X)^{-1} \Omega \vec S-\alpha \vec x.
\end{equation}
The technical condition $\lim_{N\rightarrow \infty} N\langle a_{ij}\rangle_A=c_{ij}$ applies in the case of memristive dynamics if we consider the fact that $\Omega$ a projector operator. As shown in \cite{Caravelli2019Ent}, typically, $\Omega_{ij}\approx \frac{c}{N}$ because of the condition $\Omega^2=\Omega$. As such, if we identify $A=\chi \Omega X $, since we impose $0<x_i<1$ dynamically, and $\chi<1$, then we have that $a_{ij}<\frac{\chi |c|}{N} $, which satisfies the technical condition.
  The numerical results are obtained for a number of memristors equal to $N=200$, and the of the span of $|Span(\Omega)|=N-M$ with $M=50$. For random matrices, we generate $\Omega=A^t(A A^t)^{-1} A$, with $A$ a random matrix of size $N\times M$, and $a_{ij}$ randomly distributed in $[0,1]$.

We now discuss how to apply the result above to the dynamics of memristors. Let us define $f_i(\vec x)=\sum_{j} \Omega_{ij} x_j$. It is not hard to see from the definition above that $z_i=\chi f_i(\vec x)$. Let us define $\vec f=\{ f_i(\vec x)\}$.
We then have
\begin{eqnarray}
\tilde A=\begin{pmatrix}
        f_1(\vec x) & \cdots & f_1(\vec x)\\
        f_2(\vec x) & \cdots & f_2(\vec x)\\
        \vdots & \ddots & \vdots\\
        f_N(\vec x) & \cdots & f_N(\vec x)
        \end{pmatrix}= \vec f\otimes \vec 1^t.
\end{eqnarray}

 The expansion above can be rewritten as
\begin{eqnarray}
(I-\chi \Omega X)^{-1}= I+\frac{1}{N}\frac{\chi}{1-\chi\frac{1}{N}\sum_{i=1}^N f_i(\vec x)}\vec f \otimes \vec 1^t +O(\frac{1}{N})
\end{eqnarray}
from which we obtain, if we call $ \Omega \vec S=\Omega \vec S$
\begin{equation}
    \frac{d}{dt} \vec x=\frac{1}{\beta}\Big( \Omega \vec S+\frac{\chi \frac{1}{N} \Omega \vec S\cdot \vec 1}{1-\chi \frac{1}{N} \vec f\cdot \vec 1 }\vec f\Big)-\alpha \vec x+O(\frac{1}{N})
\end{equation}
Now that we have the identity $\Omega \vec f=\vec f$ and $\Omega \vec S= s$. Thus, if we project the equation using $\Omega$, we have
\begin{equation}
    \frac{d}{dt} \vec f=\frac{1}{\beta}\Big( \Omega \vec S+\frac{\chi \frac{1}{N} \Omega \vec S\cdot \vec 1}{1-\chi \frac{1}{N} \vec f\cdot \vec 1 }\vec f\Big)-\alpha \vec f+O(\frac{1}{N}).
    \label{eq:meanfield}
\end{equation}
The asymptotic states are thus determined by the equation
\begin{eqnarray}
\frac{1}{\beta}\Big( \Omega \vec S+\frac{\chi \frac{1}{N} \Omega \vec S\cdot \vec 1}{1-\chi \frac{1}{N} \vec f\cdot \vec 1 }\vec f\Big)-\alpha \vec f=0
\end{eqnarray}
or 
\begin{eqnarray}
\vec f=\Omega \vec x=  \frac{1}{1-\frac{\chi}{\alpha \beta} \frac{\langle \Omega \vec S\rangle}{1-\chi x_{cg}}}\frac{\Omega \vec S}{\alpha \beta},
\end{eqnarray}
where we defined $\langle  \vec G\rangle=\frac{1}{N} \vec G\cdot \vec 1$. The asymptotic states
 can be obtained up to a gauge transformation $\vec x^*=\vec x^*+(I-\Omega) \vec r$ for an arbitrary $\vec r$. The equation above however cannot be solved without first finding a value for $x_{cg}.$
An equation for $x_{cg}=\frac{1}{N} \vec f\cdot \vec 1$ can be obtained by summing over all the indices on both sides of  equation (\ref{eq:meanfield}) and dividing by $N$, obtaining 
\begin{eqnarray}
\partial_t x_{cg}=\frac{1}{\beta}\Big( \langle \Omega \vec S\rangle+\frac{\chi \langle \Omega \vec S \rangle}{1-\chi x_{cg}}x_{cg}\Big)-\alpha x_{cg}+\mathcal L(\vec x)
\label{eq:meanfielddyn}
\end{eqnarray}
where $\mathcal L(\vec x)$ is an effective force we will discuss in a moment.

In order for the equilibrium points to be compatible , we thus have to solve also for the equation
\begin{eqnarray}
\frac{1}{\beta}\Big( \langle \Omega \vec S\rangle+\frac{\chi \langle \Omega \vec S \rangle}{1-\chi x_{cg}}x_{cg}\Big)-\alpha x_{cg}=0
\end{eqnarray}
which is equivalent to a mean-field equation of the form:
\begin{eqnarray}
x_{cg}=\frac{\langle \Omega \vec S\rangle}{\alpha \beta }(1+\frac{\chi x_{cg} }{1-\chi x_{cg}} )
\end{eqnarray}
whose solution is given by
\begin{eqnarray}
x_{cg}^*=\frac{1}{N}\sum_{ij} \Omega_{ij} x_j^*=\frac{\alpha \beta-\sqrt{\alpha^2\beta^2-4 \alpha \beta \chi \langle \Omega \vec S\rangle}}{2 \alpha \beta \chi}
\end{eqnarray}

By setting $\frac{\langle \Omega \vec S\rangle}{\alpha \beta}=s$ we obtain a mapping to the one dimensional case. Let us now discuss the emergence of the lack of convexity of the potential $V(x,s)$ as a function of $s$; specifically, let us analyze the situation in which the potential has a minimum or a maximum. The coarse grained variable is given, as mentioned earlier, by $x_{cg}=\frac{1}{N} \sum_{ij} \Omega_{ij} x_j$.
First, we focus on the potential on a compact support $D=[x_{min},x_{max}]$, determined via the condition $x_{min}=min_{\vec x\in[0,1]^N}(x_{cg})$  and $x_{max}=max_{\vec x\in[0,1]^N}(x_{cg})$. For $s<0$ the potential has only a local minimum (which is also an absolute minimum) at $x_{cg}=\text{max}(0,x_{min})$. For $s>0$, the potential of Eqn. (\ref{eq:potential}) can have at most two points in which its derivative in $x$ is zero, given by the values $x^{\pm}=\frac{1\pm \sqrt{1- 4 s \chi }}{2 \chi }$. 
While $x^-$ is always a local minimum, $x^+$ is always a local maximum of the potential.  However, there is an intermediate set of values, given by $s=[0,\frac{1}{4 \chi}]$, in which the potential has only a single minimum in the domain $[x_{min},x_{max}]$, as $x^+$ is outside the domain.
In this case, the local minimum $x^{-}$ is also an absolute minimum in the domain $[0,1]$. However, if $\chi$ is close enough to $1$ and $s$ large enough, we have that $x^{+}$ moves inside the domain, and thus the potential has two local minima in $[0,1]$, one at $x^{-}$ and one at $x_{max}$. If $s$ increases further, we can have that $V(x_{max})<V(x^-)$, implying that the potential is not non convex, and a barrier of height $\Delta E(s)=V_s(x^{+},\chi)- V_s(x^-,\chi)$ emerges between the the local minimum $x^{-}$ and the absolute minimum at $x_{xmax}$. 

\section*{C: Projected Dynamics and equilibrium points}
One of the advantages of using an analytical framework to study memristor dynamics is that many subtleties of nonlinear circuits become explicit in this setup.
In this regard, it is interesting to recall the fact that $\Omega^2=\Omega$. As mentioned earlier, the fact that this matrix is a projector operator is essentially a consequence of the Kirchhoff' laws. Such freedom implies a redundancy in the way in which we can effectively represent the dynamics, as investigated in earlier papers \cite{Caravelli2016ml,Caravelli2016rl,Caravelli2017b,Caravelli2019Ent}. In fact, in order to obtain the matrix $\Omega$ from the circuit, one has the liberty of choosing a particular reference spanning tree, whose details are clarified in \cite{Zegarac}. Such freedom is not hidden, but is evident from the fact that if we perform the transformation 
$\vec S^\prime=\vec S+(I-\Omega) \vec k$ for an arbitrary vector $\vec k$, the dynamics is unchanged (this is reminiscent of the ``toy" gauge freedom which is typically present in circuits). We now see that such freedom is also present in the representation of the fixed points. We can in fact write  $\vec x_j^\prime=\vec x_j+(I-\Omega) \vec k$, without effectively changing the coarse grained variable fixed point state.

Let us consider the two formulations of the dynamics, based on the fact that $\vec x=\Omega \vec x+(I-\Omega) \vec x=\vec x_c+\vec x_e$, in which $\vec x_c\cdot \vec x_e=0$ because $\Omega$ is an orthogonal projector.

Given the dynamics introduced in the bulk of the paper, we have
\begin{eqnarray}
 \frac{d\vec x}{dt}=\frac{1}{\beta}(I-\chi \Omega X)^{-1} \Omega \vec S-\alpha \vec x,
\end{eqnarray}
from which, projecting via $\Omega$ and $\Omega_B=I-\Omega$, we have, using the series expansion for the matrix expansion, that
\begin{eqnarray}
 \frac{d\vec x_c}{dt}&=&\frac{1}{\beta}(I-\chi \Omega (X_e+X_c))^{-1} \Omega \vec S-\alpha \vec x_c, \nonumber  \\
  \frac{d\vec x_e}{dt}&=&-\alpha \vec x_e.
\end{eqnarray}
It follows that since $\vec x_e$ falls exponentially to zero, the fixed points are those such that $\vec x_e=0$. Given $\vec x_c=\vec f$, the basin of attraction for the fixed points $\vec f^*$ is up to an arbitrary vector $\vec x_e=(I-\Omega)\vec k$ added to the initial conditions.
For the asymptotic fixed points emerging from the mean field dynamics, this implies that, since $x^*_{cg}$ must be such that $\frac{1}{N}\sum_{ij} \Omega_{ij} x_j^*=a$, then necessarily $\sum_j \Omega_{ij} x_j^*=x_i^*$. Thus, if $\Omega_{ij}$ has a span of $N-M$ vectors, $\vec x^*=\sum_{k=1}^{M-N} a_k \vec n_k$, where $\vec n_k$'s are the basis vectors of $\text{Span}(\Omega).$ It follows that the solutions of the problem are such that, if we call $\tilde n_k=\sum_{i} (\vec n_k)_i$, the solutions are of the form $\sum_k a_k \tilde n_k=a$.

\section*{D: Local stability analysis and pseudospectral triviality}
\label{sec:lle}%
For the single memristor dynamical system, the stability of the equilibrium is determined by the condition $-\partial_x^2 V(x^*)>0$, with $\partial_x V(x)=0$. This is equivalent to the condition
\begin{eqnarray}
 \frac{2 \alpha  \beta  \left(-\alpha  \beta +\sqrt{\alpha  \beta  (\alpha  \beta -4 s \chi )}+4 s \chi
   \right)}{\left(\sqrt{\alpha  \beta  (\alpha  \beta -4 s \chi )}-\alpha  \beta \right)^2}<0.
\end{eqnarray}
For reference, in the main text we consider the values $\alpha=\beta=1$ and $\chi=0.9$, and the dynamical instability  occurs around $s\approx 0.27777(8)$.

The dynamics of our system can be described by the differential equations:
\begin{equation}
\frac{d {\vec x}}{dt}={ \vec f}({\vec x})
\end{equation}
where ${\vec f}=[f_1, f_2 \cdots, f_S]^T$ is a set of non-linear functions  \cite{ott}. 

Since we are interested in our system escaping from the equilibrium point, we study numerically the equilibria ${\vec x*}$ which must satisfy
\begin{equation}
{\vec x}({\vec x^*})=0.
\end{equation}
Since the system can have an exponential number of fixed points, such search cannot be exaustive. However, finding these points is relatively easy and can be done numerically by initializing the dynamics randomly and letting the system evolve into one of these minima.

The local dynamics near each equilibrium point is described by the Jacobian matrix evaluated at an equilibrium point.
An equilibrium point is stable if under any infinitesimally small perturbation, $\Delta {\vec x}(0)$,
eventually decays to zero, i.e.,  $\lim_{t\rightarrow \infty} \Delta {\vec x}(t)=0$.
In the vicinity of an equilibrium point, the time evolution of a perturbation can be written as
\begin{equation}
\Delta {\vec x}(t) = e^{{ DJ}t}\Delta {\vec x}(0).
\end{equation}
Therefore, the spectrum of $DJ$ is relevant for local stability analysis.
If $\Lambda(${ M}$)$ is the set of eigenvalues of ${\bf M}$,
then the equilibrium point is stable if all eigenvalues have negative real part.
For stability near the boundaries,  the dynamical system is given by 
\begin{eqnarray}
 \frac{d}{dt} x_j=W_j(x_j) f_j(\vec x)
\end{eqnarray}
where $W_p(\vec x)$ is a window function enforcing $0\leq x\leq 1$ and $f_j(\vec x)=(\frac{1}{\beta} (I-\chi \Omega X)^{-1} \Omega \vec S-\alpha \vec x)_j$. In the following we consider the non-absorbing window function $W(x_i)=\theta(-f_i) \theta(x_i)+\theta (f_i) \theta(1-x_i)$,  where we relaxed to $\theta(a)$ to $\theta_p(a)=\frac{1}{2}+\frac{1}{2}\tanh(p a)$, for numerical stability\cite{traversa}. The numerical results shown in the paper are obtained for $p=100$, which is a good approximation of the Heaviside theta function $\theta(x)$. The full Jacobian of the dynamics is given by
\begin{eqnarray}
 DJ_{ij}^f=\partial_{x_i}(W_j(x_j) f_j(\vec x))=W_j(x_j)\partial_{x_i}( f_j(\vec x))+f_j(\vec x)\partial_{x_i} W_j(x_j).
\end{eqnarray}
The Jacobian for the dynamical system without window functions can be evaluated exactly, and is given by (using $\partial_\alpha A^{-1}=-A^{-1} (\partial_\alpha A) A^{-1}$) by ($\delta_{ij})$ is the Kronecker delta):
\begin{eqnarray}
DJ_{ji}&=&\partial_{x_i} f_j(\vec x) \nonumber \\
&=&
\frac{\chi}{\beta \alpha }\sum_{krpm} (I-\chi \Omega X)^{-1}_{jk} \Omega_{kr} \delta_{ri}  (I-\chi \Omega X)^{-1}_{rm} (\Omega \vec S)_m - \delta_{ji}\nonumber \\
&=&\frac{\chi}{\alpha\beta} P_{ji} V_i-\delta_{ji},
\end{eqnarray}
where $P=(I-\chi \Omega X)^{-1}\Omega$ and $\vec V=P\vec S$, and 
which we could use to evaluate the Local Lyapunov Exponents or study local stability. 
The local flow  divergence is obtained via the relation
\begin{eqnarray}
\delta \vec x (t_{n+1})=\big(I+DJ^f(t_n) dt\big) \delta \vec x (t_{n}).
\end{eqnarray}


The spectrum of the Jacobian at the fixed point has been studied first in \cite{Caravelli2016ml}, showing that the spectrum is real even if the matrix is not symmetric. Despite this, non-normal Jacobian matrices can in principle exhibit amplification of perturbations on a stable equilibrium and it is worth of careful scrutiny in this paper. 
Small perturbations of a stable equilibria typically decay in exponential time. In the case of non-normal matrices however, even if the eigenvalues have all real part negative, can exhibit a transient instability.
How non-normal a matrix $M$ is quantified by (any) norm of the matrix $N=(M^t M-M M^t)/2$. The Jacobian matrix of our system can be written in the form $DJ(X)=P^2 \mathcal S$, where  $P=(I-\chi \Omega X)^{-1} \Omega$ and $\mathcal S_{ij}=S_i \delta_{ij}$. It follows that $N=\frac{P^2 \mathcal S-\mathcal S(P^t)^2}{2}$.
This is not inconsistent with the notion of stability. As perturbations increase in magnitude via a transient phase, the effect on eigenvalues is more pronounced. For long times, if the system is linear, the long term dynamics is still dominated by the negativity of the real part of the spectrum.
However, in the case of non-normal matrices, the transient dynamics can be no longer consistent with the traditional picture of asymptotic stability.

For this reason we go beyond a simple spectral analysis. Generalizations of the notion of eigenvalues have been studied in the literature.
In general, the eigenvalues of ${DJ}$ can be defined:
$\Lambda({DJ}) = \{z \in \mathbb{C} : \text{det}(z I-DJ) = 0\}$, 
meaning that if $z$ is an eigenvalue of ${DJ}$ then by convention the
norm of $(z I-DJ)^{-1}$ is defined to be infinity~\cite{TrefethenEmbreeBook2005}.
But if $||(z I-DJ)^{-1}||$ is finite and very large, as is often the case with perturbed non-normal matrices, 
then the pseudospectrum of ${DJ}$ must be considered. The `$\epsilon$-pseudospectrum' is a generalization of the notion of eigenvalues which depend on a real parameter $\epsilon$, and can be defined in various equivalent ways \cite{TrefethenEmbreeBook2005}. 
We use the following definition:
\begin{equation}
\Lambda_{\epsilon}({DJ})=\{z \in \mathbb{C} : ||(z I-{DJ})^{-1}|| \geq \epsilon^{-1}\}.
\end{equation}
If a matrix is normal then its $\epsilon$-pseudospectrum (from now one we only call it `pseudospectrum') 
is somewhat trivial: it consists of closed balls of radius $\epsilon$ surrounding the original eigenvalues of $DJ$. 
In the case of non-normal matrices however, pseudospectra can be much larger and  much more intricate, which is the case we are interested in here.

Local asymptotic stability is determined in the same way for normal and non-normal matrices. 
The `spectral abscissa' of ${DJ}$ is defined as
$\alpha({DJ})=\sup_{z\in \Lambda({DJ})} \text{Re}(z)$, 
where the supremum is the largest real part of $\Lambda({DJ})$; clearly, stability is guaranteed for $\alpha({DJ}) < 0$.
If ${DJ}$ is normal, then $||e^{{DJ} t}|| = e^{\alpha({DJ}) t}$ and dynamics is determined by $\alpha(DJ)$. 
The relationship between the spectral abscissa and the system dynamics is evident from the bouns \cite{TrefethenEmbreeBook2005}:
$e^{\alpha({DJ}) t} \leq ||e^{{DJ} t}|| \leq \kappa({\vec V}) e^{\alpha({DJ}) t}$
where $\kappa({\bf V})=||{\bf V}||\cdot||{\bf V}^{-1}||$ is called the conditioning of the matrix ${\bf V}$,
which is built from the eigenvectors of ${DJ}$, and is a measure of invertibility of ${\bf V}$.
We see that for a normal dynamics, the conditioning provides an upper bound to the maximum amplification of the perturbations. Clearly, if the spectral abscissa is positive, perturbations are unstable but within the bounds defined above.
These bounds can however generalized by introducing the notion of  `$\epsilon$-pseudospectral abscissa' of ${DJ}$. This is defined as $\alpha_{\epsilon} ({DJ})=\sup_{z \in \Lambda_\epsilon({DJ})} \text{Re}(z)$, e.g. the largest real part of the spectrum of $DJ$ (for a given $\epsilon$).
The relevance of the $\epsilon$-pseudospectral abscissa is given by the fact that it provides a lower bound to 
the maximum amplification of the perturbation\cite{TrefethenEmbreeBook2005}:
\begin{equation}
\sup_{\epsilon\geq 0} \frac{\alpha_\epsilon({DJ})}{\epsilon} \leq \sup_{t \geq 0} ||e^{{DJ} t}||.
\label{Eqn:lowerbound}
\end{equation}
The quantity
\begin{equation}
f_{DJ}(\epsilon)=\frac{\alpha_\epsilon({DJ})}{\epsilon}
\label{Eqn:fM}
\end{equation}
is called the Kreiss constant of the matrix $DJ$,
and is thus relevant in order to understand the transient dynamics for short times.
Given the bound of Eqn. (\ref{Eqn:lowerbound}), it is important to look for values $\epsilon^*$, at which $f_{DJ}(\epsilon^*)=1$.
Via a visual analysis of the contours of pseudospectrum, a necessary condition of a critical value of $\epsilon^*$ is that $\epsilon^*$-contour crosses the imaginary axis.
At this point, perturbations from the equilibrium population vector can  be in principle amplified, depending on the value of the Kreiss constant.

\begin{figure}
    \centering
    \includegraphics[scale=0.12]{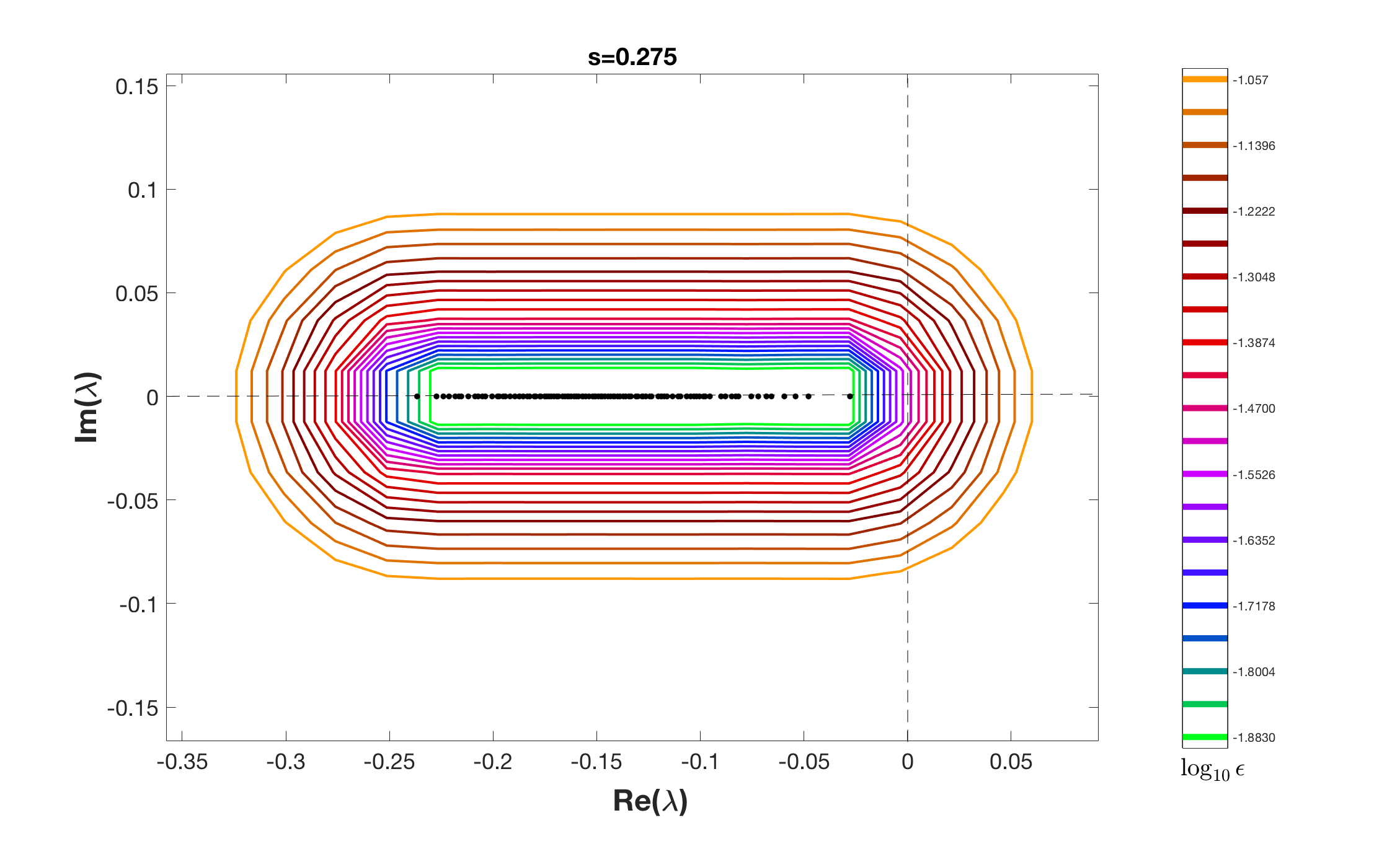}    \includegraphics[scale=0.19]{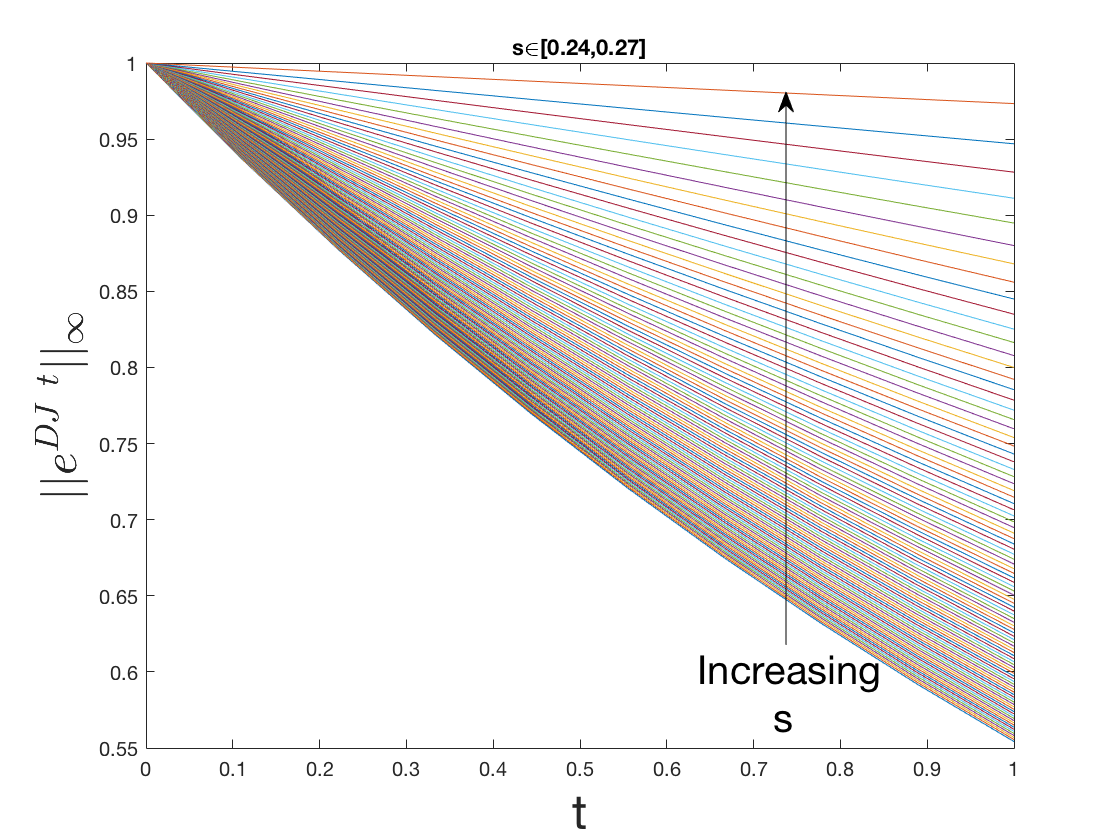}

    \caption{Left: $\epsilon$-Pseudospectrum of the Jacobian matrix for $\epsilon=[10^{-2},10^{-1}]$.  Right: behavior of the norm $||e^{DJ\ t}||_{\infty}$ as a function of time for $s\in [0.24,0.27]$.}
    \label{fig:nonnorm1}
\end{figure}

In order to study the non-normality, we looked at the Jacobian matrix evaluated at the equilibrium points for the dynamics obtained numerically via Monte Carlo.
In Fig. \ref{fig:nonnorm1} (left) we can observe the bulk of $\epsilon$-pseudospectrum of the Jacobian matrix $DJ$, fixed $\Omega$ as in the main text, and with $\alpha=\beta=1$ and $\chi=0.9$. However, such non-normality is not enough to explain the transient instability we observe. A careful numerical analysis of the Kreiss constant for varying $s$ gave us a maximum value of the Kreiss matrix of $K=sup_\epsilon F_{DK}(\epsilon)=0.7$, thus less than 1, the critical value for the amplification. 
We thus directly studied the norm of $e^{DJ t}$ in Fig. \ref{fig:nonnorm1} (right), not observing any amplification phenomenon up to the instability threshold for the matrix. This implies that $\sup_{\epsilon} \frac{\alpha_\epsilon(DJ)}{\epsilon}\leq 1$ and that transient instability is not the cause of the escape from the minimum.

\section*{E: Basins of attraction}  \label{sec:bas}

\begin{figure}
    \centering
        \includegraphics[scale=0.17]{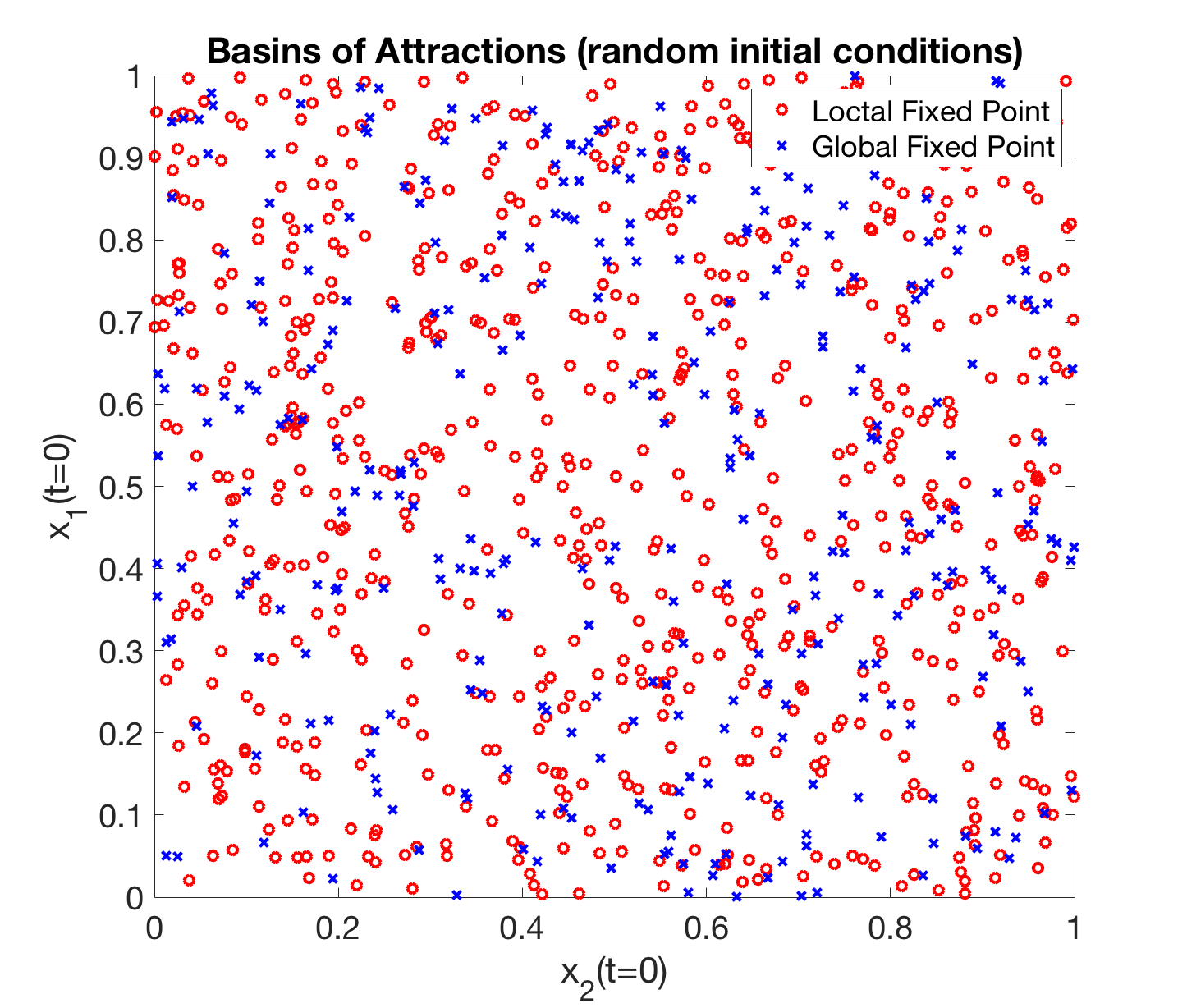}
        \includegraphics[scale=0.17]{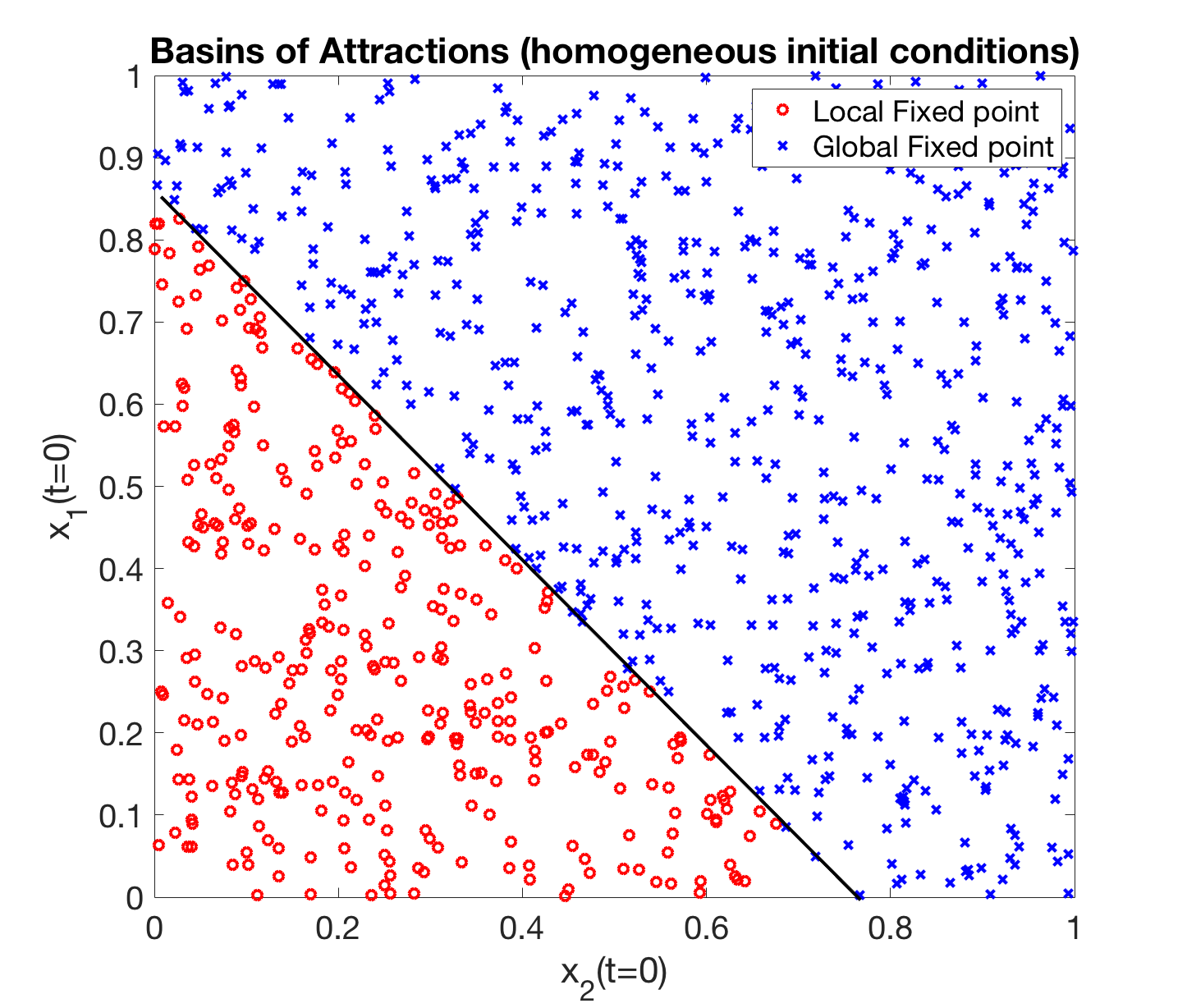}
    \caption{Projected basin of attraction for the coarse grained asymptotic states according to the two controlled numerical experiments, as described in the text, in order to visualize the basins of attractions. On the left, we see the completely randomize initial condition, in which we see a complete mixing in the initial conditions for $x_1$ and $x_2$ as a function of the asymptotic states. On the right, we see that there is a neat separation between the two asymptotic states via a separatrix bordering two basins of attractions.}
    \label{fig:basins}
\end{figure}

First of all, let us first mention, as shown in the main text, that the tunneling probability depends on the number of dynamical variables. In Fig. \ref{fig:tunnellingPt} we see that the larger the height of the barrier ($\Delta E$ decreases with $s$), the larger the number of variables one needs to tunnel through.
In order to obtain a glimpse of the structure of the basins of attractions and the role of the dimensionality in reaching  the two asymptotic states we perform the following numerical experiment.
 Given a random instance of $\Omega$ obtained as $\Omega=A^t(AA^t)^{-1} A$, for $A$ a random matrix (with flat probability on $[0,1]$) of size $N_c\times N$ ($N=200$, $N_c=100$)  we fix $\chi$ and $s=\Omega \vec S$ in a region in which we know there is a mixing between the local minimum and global minimum of the mean field potential potential $V(x,\chi)$ in $x\in[0,1]$. Specifically, we study the basins of attraction of the system near the mean field situation (e.g. a single memristor), e.g. when many of the variables $x_i$ are initialized in the same value.
 We consider this analysis in the regime reported in the main text,  for the values $\chi=0.9$ and $s\approx 0.22$ which is close to the boundary between the two minima. We then perform the following two experiments. In the first, we choose $x_i(0)$ randomly, and then follow the trajectory and mark it as local or global minimum asymptotic state. Since $\Omega_{ij}$ is random and there is no preference between the variables, we choose to plot initial conditions of the variables $x_1$ and $x_2$ and mark them depending on which asymptotic state they reached. In the second experiment instead we choose $x_{3}\cdots x_{N}=\tilde x$, and $x_1$ and $x_2$ still at random, following the procedure of the first experiment and following the trajectory. The value of $\tilde x$ is an extra parameter and has to be obtained such that system ends in both minima. A value for $\tilde x$ which we found to be working well for random $\Omega$ is $\tilde x\approx .5135$. The results are shown in Fig. \ref{fig:basins}, in which we see a complete mixing in the first experiment between the asymptotic states, and a complete separation via a separatrix in the second experiment. The separatrix moves with $\tilde x$, and for larger values the system it tends to move towards the global minimum; for smaller values towards the local minimum. Intuitively, this implies that for random initial conditions, for every possible asymptotic state there is a initial condition nearby to the global minimum.

\begin{figure}
    \centering
    \includegraphics[scale=0.18]{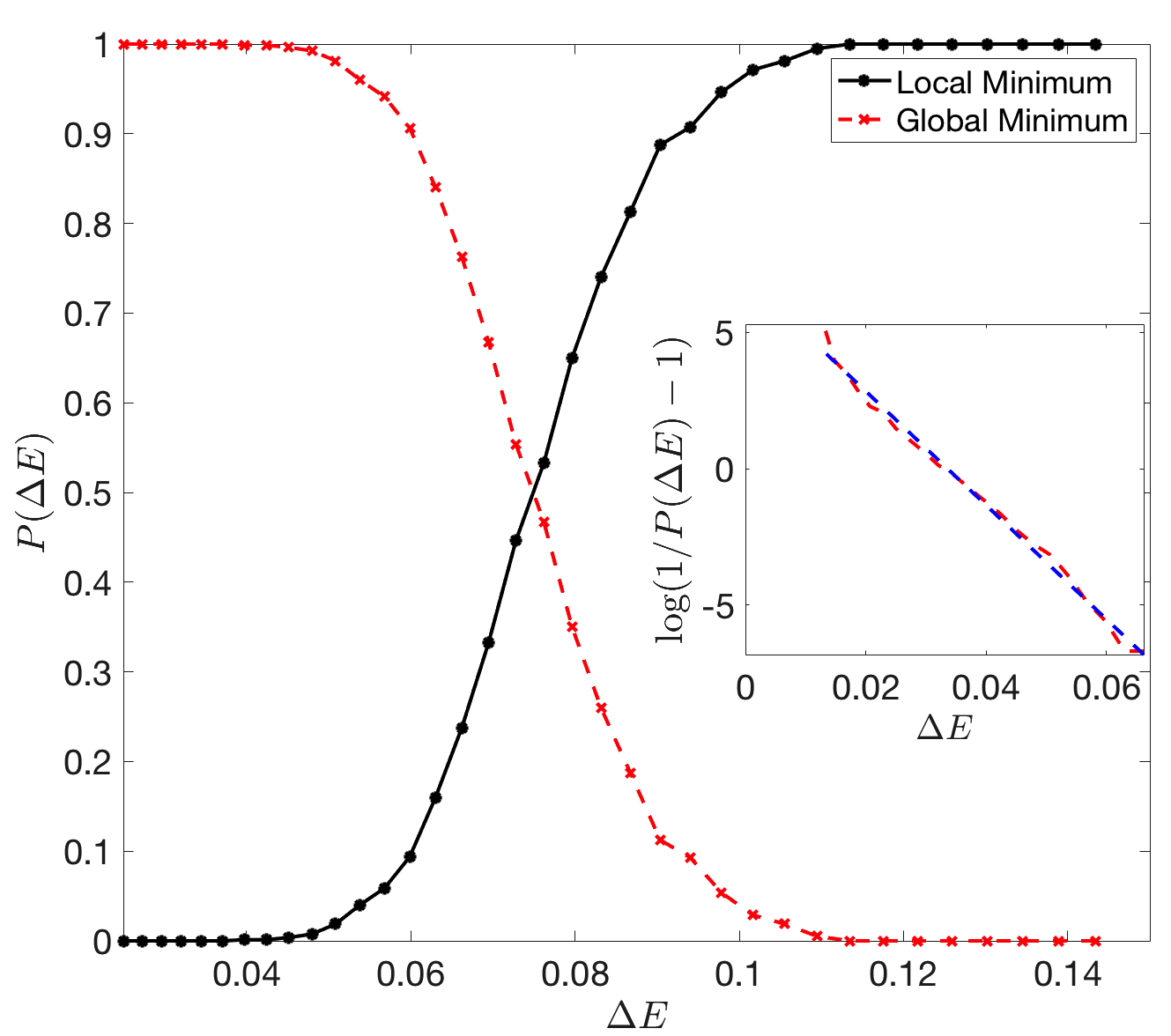}
    \caption{Distributions $P(x^1_{cg})$ and $P(x^2_{cg})$ as a function of $\Delta E$ of the potential. Inset: Fit of the Fermi distribution with a chemical potential.}
    \label{fig:effermi}
\end{figure}
 A more careful analysis has been done for random initial conditions. We studied via Monte Carlo sampling the probability distribution, given a random initial condition, of the system being in one or the other local minimum, showing that it can be fit by a Fermi-like distribution depending on the energy barrier $\Delta E$. Let $P(x; s,\chi)$ be the probability distribution as a function of the height of the barrier $\Delta E(s)=E_{max}-E_{local\ min}$ from the local minimum. We obtained such function numerically, averaging over 100 Monte Carlo simulations for each value of $s$, and shown in Fig. \ref{fig:effermi}. We  fit $P(\Delta E)$ with a function of the form $P(x; s,\chi)=1/(1-e^{-{\Delta E(s,\chi)-\mu}/{T_{eff}}})$, where we obtained numerically the value $T_{eff}=0.0048$, and $\mu=0.036$. These results are shown in Fig. \ref{fig:effermi}. We see that the transition probabilities of Fig. \ref{fig:effermi} are consistent with the sigmoidal probabilities described in the main text.

\end{document}